\documentclass[aps,12pt,a4paper]{revtex4-2}
\usepackage{epsfig}
\usepackage{graphicx}
\usepackage{amsmath,amssymb,color}
\usepackage[english]{babel}
\usepackage{natbib}
\usepackage{float}
\usepackage{subfigure}
\usepackage{enumitem}

\parskip=\medskipamount

\newcommand{\eq}[1]{(\ref{#1})}
\newcommand{\fig}[1]{Fig.~\ref{#1}}

\newcommand{\be}{\begin{equation}}
\newcommand{\ee}{\end{equation}}
\newcommand{\beq}{\begin{equation}}
\newcommand{\eeq}{\end{equation}}
\newcommand\disp{\displaystyle}

\newcommand{\eps}{\varepsilon}

\begin{document}

\title{Three faces of random walks in a hyperbolic domain: BKT, Lifshitz tails, and KPZ}

\author{Daniil Fedotov$^{1}$, Sergei Nechaev$^{2,3}$}

\affiliation{$^1$UFR Sciences, Universit\'e Paris Saclay, 91405 Orsay Cedex, France \\
$^2$LPTMS, CNRS -- Universit\'e Paris Saclay, 91405 Orsay Cedex, France \\
$^3$ BIMSA, Yanqi Lake, Huairou District, Beijing 101408, China}

\begin{abstract}

We show that continuous random walks (diffusion) in the Poincar\'{e} upper half-plane $\mathbb{H}^2 = 
\{(x,y)| y>0\}$ provide a unified description of three seemingly unrelated phenomena: (i) the non-analytic divergence of the correlation length at the Berezinskii--Kosterlitz--Thouless (BKT) transition; (ii) the emergence of a Lifshitz tail (LT) in the 1D statistics of rare events; and (iii) the appearance of Kardar--Parisi--Zhang (KPZ) scaling in the fluctuations of stretched random walks constrained above an impermeable disc. We adapt the renormalization-group (RG) analysis originally developed for the Schrödinger equation with mixed boundary conditions in a 2D conformally invariant potential to the spectral problem in $\mathbb{H}^2$, and identify the conditions under which BKT-type behavior emerges. We show that the same RG equations can be interpreted in terms of the statistics of random walks in the large-deviation regime through an instanton approach, leading to Lifshitz tails (LT) in the associated spectral problem. We further derive KPZ-type scaling for the fluctuations of stretched random walks near the boundary of a disc in $\mathbb{R}^2$ and emphasize the similarity of this behavior with the boundary fluctuations of paths in $\mathbb{H}^2$ analyzed using a WKB approach.

\end{abstract}

\date{\today}

\maketitle

\section{Introduction: Background and Objectives}
\label{sect:01}

Continuous random walks on hyperbolic manifolds provide a powerful framework for understanding diffusion and fluctuation phenomena in negatively curved geometries, where the volume grows exponentially with distance and heat kernels exhibit characteristic exponential decay with respect to the geodesic distance \cite{DaviesMandouvalos1988}. Throughout this paper, we consider the hyperbolic manifold realized as the Poincar\'{e} upper half-plane, $\mathbb{H}^2 = {(x,y), | -\infty < x < \infty, y > 0}$. Diffusion on this manifold is governed by the Beltrami-Laplace operator, and the corresponding stochastic process, hyperbolic Brownian motion, has been studied through its transition semigroup and heat kernel \cite{Gruet1996}. The non-Abelian symmetry structure of hyperbolic space also provides a natural connection between stochastic diffusion and the geometry of its isometry group. Furthermore, hyperbolic Brownian motion is closely related to exponential functionals of ordinary Brownian motion and to Bessel processes \cite{Vakeroudis2012, Yor1992}. These representations provide a probabilistic bridge between hyperbolic and Euclidean diffusions. They also lead naturally to non-Gaussian fluctuation phenomena associated with exponential Brownian functionals and to large deviation principles for the radial component of hyperbolic Brownian motion \cite{Hirao2011}. In this sense, the Beltrami-Laplace operator provides a useful link between the geometry of negatively curved spaces, Bessel-type Euclidean diffusion, and the rare-event statistics of hyperbolic stochastic processes.

\subsection{Diffusion in 2D Euclidean plane $\mathbb{R}^2$: Naive scaling estimates}
\label{sect:01a}

Let us begin with the textbook example of the continuous diffusion process in the 2D Euclidean plane $\mathbb{R}^2$. The probability distribution $P(x,y,t)$ satisfies the diffusion equation
\be
\partial_t P(x,y,t) = \tfrac{a^2}{4} \left(\partial^2_{xx}+\partial^2_{yy} \right)P(x,y,t)
\label{eq:01}
\ee
where $t$ ($t\gg 1$) is the dimensionless diffusion time, and ${\cal D}=\tfrac{a^2}{4}$ is the two--dimensional diffusion coefficient. The diffusion process begins at $(x_0,y_0)$ and terminates at $(x,y)$ in the unbounded space $\mathbb{R}^2$. In terms of random walks, $t$ can be interpreted as a number of steps in the walk (considered as a continuous variable), and $a$ -- as the typical length of the elementary step, such that $L=ta$ is the total length of the walk. Throughout the work we will not distinguish between the "diffusion" and the "random walk" \cite{MortersPeres2010, Lawler1991}.

Let $P(q_x,q_y,s)$ be the Laplace--Fourier image of $P(x,y,t)$:
\be
P(q_x,q_y,s) =\int_0^{\infty} e^{-st} dt \int_{-\infty}^{\infty}\int_{-\infty}^{\infty}e^{i (q_x x+q_y y)} P(x,y,t)dx dy
\label{eq:02}
\ee
Applying \eq{eq:02} to \eq{eq:01} we get the equation for the Green function in the Fourier space:
\be
P(q_x,q_y,s) = \frac{1}{s+\tfrac{a^2}{4}(q_x^2+q_y^2)}
\label{eq:03}
\ee
which after the inverse Fourier transform reads:
\be
P(x,y,s) =\frac{1}{\pi^2} \int_{-\infty}^{\infty}\int_{-\infty}^{\infty}\frac{e^{-i (q_x x+q_y y)}}{4s+a^2(q_x^2+q_y^2)} dq_x dq_y = \frac{2}{\pi a^2} K_0
\left(\frac{2r}{a}\sqrt{s}\right)
\label{eq:04}
\ee
where $K_0(...)$ is the modified Bessel function and $r=\sqrt{(x-x_0)^2+(y-y_0)^2}$. Equation \eq{eq:04} and its extension to the diffusion in the hyperbolic geometry are the starting point of our forthcoming discussions. The Euclidean heat kernel and its associated Green function provide the standard starting point for this construction \cite{Grigoryan2009}.

The main statistical characteristics of the diffusion process in $\mathbb{R}^2$ one can qualitatively deduce on the basis of \eq{eq:04}. In particular, we are interested in two questions:
\begin{itemize}
\item[(i)] What is the typical spread of the diffusion after time $t$ in the $(x,y)$--plane;
\item[(ii)] What is the typical probability density for the diffusion process in $\mathbb{R}^2$ to return to the starting point (i.e. to make a "Brownian bridge") at time $t$.
\end{itemize}
The recurrence properties of two-dimensional Brownian motion and the corresponding return behavior are standard consequences of the planar heat kernel and potential theory \cite{MortersPeres2010}.

From the structure of the Laplace--Fourier transformed kernel \eq{eq:04} in $\mathbb{R}^2$ one sees that the argument of $K_0(...)$ changes behavior at
\be
\frac{2r}{a}\sqrt{s_{\rm c}} \sim 1 \quad\rightarrow\quad s_{\rm c}\sim \frac{a^2}{4r^2}.
\label{eq:05}
\ee
Namely,
\begin{itemize}
\item For $s\ll s_{\rm c}$, the argument of $K_0(...)$ is small, and $P(r,s)$ behaves logarithmically.
\item For $s\gg s_{\rm c}$, the argument of $K_0(...)$ is large, and $P(r,s)$ decays exponentially in $\sqrt{s}$;
\end{itemize}
The scale $s_{\rm c}$ defines inverse diffusive times at which the Green function $P(r,s)$ starts "feeling" the whole available space region $r$ in $\mathbb{R}^2$. Namely, in probabilistic terms, $s_{\rm c}^{-1}$ is the typical diffusive time needed to reach distance $r$. When performing the inverse Laplace transform, the relevant values of $s$ are concentrated around $s^{}\sim t^{-1}$. If $s^{}\approx s_{\rm c}$, then the crossover of the Green function \eq{eq:04} coincides with the region that provides the dominant contribution in the integral for the inverse Laplace transform, and this gives precisely the diffusive scaling regime $r\sim a\sqrt{N}$, which corresponds to the majority of trajectories in the ensemble of paths. By contrast, the characteristic value $s_{\rm c}$ itself does not specify the scale of the majority of trajectories in the ensemble.

If the Green function $P(r,s)$ is concentrated within a disk of area $A(s)=\pi r^2(s)$, then a crude estimate of the probability density, $\rho(s)$, per unit area associated with the scale $s=s_{\rm c}$, is given by the inverse area of a disc. Namely, $\rho(s_{\rm c}) \sim 1/A(s_{\rm c})$, and in the grand canonical ensemble we have:
\be
\rho(s_{\rm c}) \sim \frac{1}{\pi r^2(s_{\rm c})} \sim \frac{4s_{\rm c}}{\pi a^2}.
\label{eq:06}
\ee
Again, if $s^{}\approx s_{\rm c}$, where $s^{}= t^{-1}$, then for the probability density, $\rho(t)$ in the microcanonical ensemble, we get
\be
\rho(t) \sim \frac{4}{\pi ta^2}.
\label{eq:07}
\ee
Supposing the uniformity of the probability density within a disc of radius $r$ (which is certainly a very crude approximation), the function $\rho(t)$ provides the scaling for the probability density to make a Brownian bridge at time $t$ in $\mathbb{R}^2$.

\subsection{Diffusion in 2D Poincar\'{e} half-plane $\mathbb{H}^2$: Scaling estimates for statistical observables}
\label{sect:01b}

Let us begin by recalling the solution of the 2D diffusion in the Poincar\'{e} upper half-plane $\mathbb{H}^2 = \{(x,y)\,|-\infty<x<\infty,\, y>0\}$. The surface $\mathbb{H}^2$ has the constant negative curvature $\varkappa = -1$ with the metric element $d\sigma^2=y^{-2}(dx^2+dy^2)$. The hyperbolic distance (the dimensionless geodesic length), $\ell$,  in $\mathbb{H}^2$ between points $(x,y)$ and $(x_0,y_0)$ is defined as follows
\be
\cosh \ell = 1 + \frac{(x-x_0)^2 + (y-y_0)^2}{2\,y\,y_0}.
\label{eq:08}
\ee
The diffusion equation extending \eq{eq:01} to the hyperbolic geometry $\mathbb{H}^2$ reads
\be
\partial_t Q(x,y,t) = \tfrac{a^2}{4} y^2 \left(\partial^2_{xx}+\partial^2_{yy} \right)Q(x,y,t) 
\label{eq:09}
\ee
To make $t$ dimensionless as in Euclidean case, we set in \eq{eq:09} (and everywhere below) $a=1$.

Consider the Cauchy problem in unbounded domain $\mathbb{H}^2$ with the initial condition $Q(x,y,0)=y^2\delta_{x-x_0} \delta_{y-y_0}$. After the Laplace transform, $Q(x,y,s) = \int_0^\infty Q(x,y,t) e^{-st} \, dt$, the solution of \eq{eq:09} can be written in terms of the geodesic distance, $\ell$, and the Laplace image, $s$:
\be
Q(\ell, s) = \frac{\sqrt{2}}{\pi y_0^2} \int_{\ell}^{\infty} \frac{e^{-\nu(s) \tau }}{\sqrt{\cosh \tau - \cosh \ell}} \, d\tau\, \qquad \nu(s) = \sqrt{\frac{1}{4} + 4s}
\label{eq:10}
\ee
where the dimensionless parameter $s$ (as in \eq{eq:04}) is the inverse diffusive time and the spatial geodesic distance $\ell = d_{\mathbb{H}^2}((x,y), (x_0,y_0))$ is defined in \eq{eq:08}. Taking into account that at $\ell \gg 1$ (and hence at $y\to 0$) one gets
\be
\cosh \ell \sim \frac{(x-x_0)^2 + y_0^2}{2 y y_0} \sim \frac{C}{y} \quad \xrightarrow[y\to 0]{} \quad \ell \sim \ln \frac{2C}{y} \sim -\ln y + \mathrm{const}, 
\label{eq:10a}
\ee
we can find the leading exponential behavior of the Laplace-transformed Green function $Q(x,y,s)\equiv Q(\ell, s)$ at $\ell \gg 1$ up to the power-law corrections:
\be
Q(\ell, s)\Big|_{\ell \gg 1} \sim \frac{2}{\sqrt{\pi}\, y_0^2} \frac{\Gamma\left(\nu(s)+\tfrac{1}{2}\right)}{\Gamma\left(\nu(s) + 1\right)} \, e^{-\left(\nu(s)+\tfrac{1}{2}\right) \ell} 
\label{eq:12}
\ee
where $\Gamma(...)$ is the Gamma-function. Recall that $Q(\ell,s)$ is the Green function of an ensemble of random walks in $\mathbb{H}^2$ starting at $(x_0,y_0)$ and reaching a \emph{specific} point located at the geodesic distance $\ell$ from initial point \cite{DaviesMandouvalos1988, AnkerJi1999}. 

Our goal is to derive the estimate similar to \eq{eq:05}, where $r$ is replaced by the geodesic distance $\ell$ in $\mathbb{H}^2$. Proceeding as in \eq{eq:05} we could try to separate the typical and exponentially suppressed states in \eq{eq:12} by implying the condition $\left(\nu(s)+\tfrac{1}{2}\right) \ell\sim 1$. However due to the presence of $\tfrac{1}{2}$ in the exponent of \eq{eq:12} and since $\nu(s)\ge 0$, the condition $\left(\nu(s)+\tfrac{1}{2}\right) \ell\sim 1$ does not provide any nontrivial scaling dependence for $\ell(s)$. So, one has to eliminate $\tfrac{1}{2}$ in the exponent. As we shall see, this is feasible if we restrict our consideration to the ensemble of Brownian bridges in $\mathbb{H}^2$. 

Let $W(\ell,s)$ be the \emph{conditional} distribution of random paths in $\mathbb{H}^2$ to reach the geodesic distance $\ell$ during some diffusive time, $t_{\ell}\equiv s^{-1}_{\ell}$ under condition that at the very last step (i.e. at diffusive time $t=s^{-1}$) the path returns to the starting point. In other words, we are interested in the typical span, $\ell$, of the Brownian bridge (the path with coinciding ends). Let us emphasize that the very fact of the Brownian bridge condition pushes the problem to the large deviation regime because the spontaneous closing of ends in the hyperbolic space is exponentially suppressed due to the curvature-induced drift. It is known that under the Brownian bridge condition we "kill" the influence of the drift, however leave the hyperbolic geometry of the target space unchanged \cite{nech-sin, bougerol}. 

To estimate $W(\ell,s)$ we split the path into two independent parts of equal diffusive times, $t_{\ell} = t/2 = 1/(2s)$. The probability distribution $W(\ell,s)$ is the product of these two parts under the condition that \emph{they meet at one point somewhere at the geodesic distance $\ell\gg 1$}. Recall that the circumference $C(\ell)$ and the area $A(\ell)$ of the hyperbolic disc bounded by geodesic length $\ell$ are: 
\be
C(\ell)=2\pi \sinh \ell, \quad A(\ell) = 4\pi \sinh^2 \frac{\ell}{2}
\label{eq:13}
\ee
Thus, we can estimate the large-$\ell$ conditional distribution $W(\ell,s)$ of the Brownian bridge as follows:
\be
W(\ell,s) \sim \left\{\begin{array}{ll} C(\ell) \\ A(\ell) \end{array}\right\}  Q^2(\ell,2s) \bigg|_{\ell\gg 1} \sim \mathcal{N}_{C,A}\;  e^{-\big(2\nu(2s) +1\big)\ell} e^{\ell} \sim e^{-2\nu(2s) \ell} = e^{-2\ell \sqrt{8s +\tfrac{1}{4}}}
\label{eq:14}
\ee
where the numerical coefficients $\mathcal{N}_{C,A}$ depend on whether we consider the circumference ($C$) or the area ($A$) of the hyperbolic disc, but in both cases the scaling dependence remains unchanged. As one can see, the term $e^{-\ell}$ coming from the square of the $Q$-function exactly compensates the growth of the circumference (or area) of the hyperbolic domain, $e^{\ell}$.

The term $\tfrac{1}{4}$ in \eq{eq:14} is the characteristic spectral-boundary contribution from the hyperbolic Laplacian: the continuous spectrum of $-\Delta_{\mathbb H^2}$ starts at $\tfrac{1}{4}$ \cite{DaviesMandouvalos1988, AnkerJi1999}. To compensate $\tfrac{1}{4}$ we introduce the constant drift in the hyperbolic random walk through a shift $s\to s^*$ (i.e. consider "stretched" paths). Defining $s^*$ by the condition
\be
\tfrac{1}{4}+8s\equiv 8s^* \quad \rightarrow \quad s\to s^* = s + \tfrac{1}{32}
\label{eq:15}
\ee
we absorb the curvature contribution by $s^*$ such that the relevant inverse length becomes $\nu(2s^*)= \sqrt{8s +\tfrac{1}{4}} = \sqrt{8s^*}$. Rewriting \eq{eq:14} in terms of $\nu(2s^*)$ we represent the large-geodesic distribution \eq{eq:14} as
\be
W(\ell,s^*)\sim e^{-4\sqrt{2}\,\ell\sqrt{s^*}}
\label{eq:15a}
\ee

Now we can proceed as in \eq{eq:05}. When $\ell\sqrt{s^*}\ll 1$, the probability is not exponentially suppressed. Conversely, for $\ell\sqrt{s^*}\gg 1$, the probability becomes exponentially small, signaling the emergence of the large-deviation regime. The crossover between these regimes is determined by $\ell_{\rm c}\sqrt{s^*_{\rm c}}\sim 1$, which sets the typical dimensionless geodesic length scale, $\ell_{\rm c}$:
\be
\ell_{\rm c}\sim\frac{1}{\sqrt{s^*_{\rm c}}}
\label{eq:15b}
\ee
Physically, this relation means that $1/\sqrt{s^*_{\rm c}}$ defines the characteristic large-geodesic length scale associated with approaching the shifted spectral boundary by the random walk with the drift. As $s^*_{\rm c}\to 0^+$, the characteristic geodesic diverges. At large $\ell_{\rm c}$ for the circumference and the area one has from \eq{eq:13}: $C(\ell_{\rm c})\sim A(\ell_{\rm c}) \sim e^{\ell_{\rm c}}$. Hence a physical correlation length, $\xi_{\rm c}$, measured by the hyperbolic circumference, $C(\ell_{\rm c})$, or area, $A(\ell_{\rm c})$, grows exponentially with the geodesic radius $\ell_{\rm c}$,
\be
\xi_{\rm c}\sim e^{\ell_{\rm c}}.
\label{eq:16}
\ee
Combining this geometric relation with the spectral-edge scaling, we get
\be
\xi_{\rm c}\sim e^{\frac{b}{\sqrt{s^*_{\rm c}}}}, \qquad s^*_{\rm c}\to 0^+
\label{eq:17}
\ee
where $b$ is a nonuniversal model-dependent constant. This is the characteristic Berezinskii-Kosterlitz-Thouless (BKT) essential singularity resulting from the combination of: (i) the square-root behavior at the crossover scale from typical to large-deviation behavior of biased random walks at the spectral boundary, and (ii) the exponential growth of length and area with geodesic distance in $\mathbb H^2$. In Section \ref{sect:02} we support the scaling \eq{eq:17} by RG computations which show that the divergence of a correlation length is actually the BKT transition \cite{bkt}. 

The probability to find the midpoint of a Brownian bridge controlled by the inverse diffusion time, $s^*_{\rm  c}$, in some \emph{specific} point at the boundary $C(s^*_{\rm c})$ or within the area $A(s^*_{\rm c})$ in $\mathbb{H}^2$, can be roughly estimated as 
\be
\mu(s^*_{\rm c}) \sim \frac{1}{C(s^*_{\rm c})}\sim \frac{1}{A(s^*_{\rm c})} \sim \xi^{-1}_{\rm c} \sim e^{-\frac{b}{\sqrt{s^*_{\rm c}}}}
\label{eq:18}
\ee
Comparing the probabilities $\rho(s_{\rm c})$ in the Euclidean geometry \eq{eq:06} and $\mu(s_{\rm c})$ in the hyperbolic one \eq{eq:18} one claims that the distribution of $\mu(s_{\rm c})$ is the manifestation of statistics of rare events. We intend to interpret this behavior in Sections \ref{sect:03} and \ref{sect:04} as the Lifshitz tail (LT) \cite{lifshitz} emerging in a probability distribution of "stretched" random walk, where the external disorder, typical for standard LT--behavior is replaced by the boundary--dependent geometry.

The following remark is noteworthy. We have identified the scaling dependence in \eq{eq:15b} as a clear threshold separating two distinct asymptotic regimes of the Brownian bridge: typical and exponentially rare. However, considering the Brownian bridge itself is not obligatory. Any other characteristic distribution function whose asymptotic behavior exhibits a separation between distinct regimes under the condition $\nu(s)\ell \sim 1$ could be used instead. The advantage of the Brownian bridge is that it provides the simplest possible framework for obtaining non-rigorous, yet physically intuitive, scaling estimates. In Section \ref{sect:02}, we revisit this question more rigorously by considering the partition function of open paths confined to a bounded domain in $\mathbb{H}^2$. 

\subsection{Objectives of the work}
\label{sect:01c}

\begin{figure}[ht]
\centering
\includegraphics[width=0.8\linewidth]{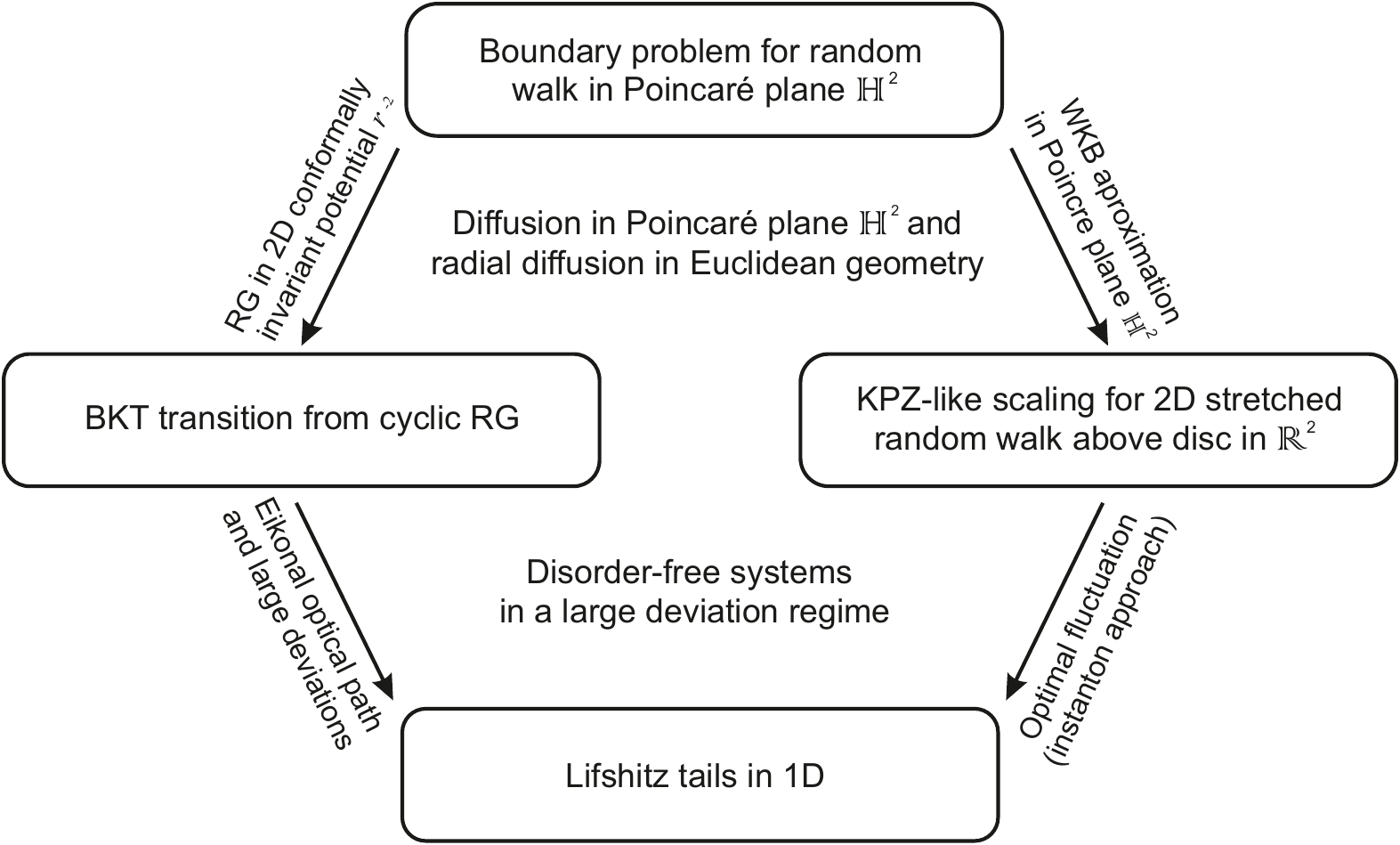}
\caption{Flowchart: Three faces of random walks in $\mathbb{H}^2$.}
\label{fig:01}
\end{figure}

We aim to demonstrate that random walks in the hyperbolic half-plane $\mathbb{H}^2$ provide a framework for a unified description of three seemingly unrelated phenomena:
\begin{itemize}
\item[(i)] The non-analytic divergence of the correlation length at the Berezinskii--Kosterlitz--Thouless (BKT) phase transition \cite{bkt};
\item[(ii)] The emergence of Lifshitz tails (LT) in one-dimensional statistics of a disorder-free system in a large deviation regime \cite{lifshitz};
\item[(iii)] The appearance of the Kardar--Parisi--Zhang (KPZ)--like scaling \cite{kpz} in the fluctuation behavior of random walks stretched above an impermeable disc in 2D Euclidean plane \cite{ferrari,nech-pol-val,fedotov-nech}, which can also be viewed as a WKB asymptotics in a bounded $\mathbb{H}^2$ domain.
\end{itemize}
Our discussion combines qualitative scaling arguments with more rigorous analytic derivations supported by numerical computations. The corresponding flowchart of the topics discussed throughout the work is schematically depicted in \fig{fig:01}.

The paper is structured as follows: in Section \ref{sect:02} we adapt the renormalization group (RG) equations initially derived for the description of the Efimov effect in a 2D conformally invariant potential \cite{kaplan, gorsky-RG, akkermans} to diffusion in $\mathbb{H}^2$. This enables us to interpret \eq{eq:17} as the divergence of the correlation length near the BKT transition at $s^* \to 0^+$, where $s^*$ plays the role of the temperature in the vicinity of the critical point \cite{hyp-efimov1, hyp-efimov2}. In Section \ref{sect:03} we discuss the manifestation of the Lifshitz tail in a deterministic large-deviation landscape in the hyperbolic plane obtained by a WKB-type approach, which mimics the rare-event survival probability of a 1D diffusion in an ensemble of traps with a Poissonian distribution on a line \cite{balagurov, donsker}. This provides support for a naive derivation of the Lifshitz tail in \eq{eq:18}. In Section \ref{sect:04} we consider the KPZ-type behavior of fluctuations and survival probability of "stretched diffusion" near the impermeable boundaries in $\mathbb{R}^2$, and show its equivalence to the solution of diffusion equation in $\mathbb{H}^2$ in WKB approximation, without any disorder \cite{fedotov-nech}. In Section \ref{sect:05} we summarize the obtained results. Appendix \ref{sect:app1} explains why the observed scaling could be attributed to the "KPZ"--like universality class, while Appendix \ref{sect:app2} provides details of the analytical derivation of the Airy-type distribution \eq{eq:sc03}.

\section{Face 1: Spectral problem in $\mathbb{H}^2$ and BKT critical scaling in conformally invariant potential}
\label{sect:02}

We begin with the diffusion equation in the Poincar\'e domain, $\mathbb{H}^2$, equipped with mixed (Robin) boundary condition. Diffusion on the hyperbolic plane and the associated heat kernel are well-established -- see, for example \cite{Matsumoto2007}:
\be
\begin{cases}
\disp \partial_t Q(x,y,t) = \tfrac{a^2}{4} y^2 \left(\partial^2_{xx}+\partial^2_{yy} \right)Q(x,y,t) \medskip \\
\disp \left.\partial_y Q(x,y,t)\right|_{y=y_0} = \left(g+\tfrac{1}{2}\right)y_0^{-1}\, Q(x,y=y_0,t)
\end{cases}
\label{eq:20}
\ee
where $g$ is a constant, and as discussed at length following \eq{eq:09}, we set $a=1$.

Since we have a boundary problem, for which the discrete spectrum is expected, it is more suitable to solve the problem separating the variables, rather than by using Laplace-Fourier transform. Seeking a solution of \eq{eq:20} in the form $Q(x,y,t) = e^{-st-i\varkappa x}Q(y)$, we arrive at the following spectral problem for $Q(y)$ in $\mathbb{H}^2$:
\be
\begin{cases}
\disp -4s Q(y)=y^2\left(d^2_{yy}-\varkappa^2\right)Q(y), \qquad y\ge y_0, \medskip \\ \disp d_y Q(y)\Big|_{y=y_0} = \left(g+\tfrac{1}{2}\right)y_0^{-1}\, Q(y=y_0)
\end{cases}
\label{eq:23}
\ee
By making use of the substitution 
\be
Q(y,s,\varkappa) = y^{1/2} P(y,s,\varkappa)
\label{eq:24}
\ee
one formally maps the spectral problem \eq{eq:23} in $\mathbb{H}^2$ onto the one in radial framing in $\mathbb{R}^2$, where $y$ stands for radial coordinate: 
\be
\begin{cases}
\disp \varkappa^2 P(y) = \left(d^2_{yy}+y^{-1}d_y\right) P(y) + \left(4s-\tfrac{1}{4}\right)y^{-2}P(y)=0 \medskip \\ \disp d_y P(y)\Big|_{y=y_0} = g\,y_0^{-1} P(y=y_0)
\end{cases}
\label{eq:25}
\ee
The effective potential $V_{\text{eff}}(y) = \left(4s - \tfrac{1}{4}\right)y^{-2}$ in \eq{eq:25} is attractive whenever $s>\tfrac{1}{16}$ and the system supports a discrete spectrum of localized bound states. The attractive inverse-square potential is a classic example of a scale-invariant quantum problem exhibiting a nontrivial renormalization-group structure and, in the supercritical regime, limit-cycle behavior \cite{Case1950, BawinCoon2003, BraatenPhillips2004}. Let us introduce a shifted value $\sigma = 4s - \tfrac{1}{4}$. Shifting $s$ means tuning the "extension" of random paths: in the regime $s > s^*=\tfrac{1}{16}$ one effectively applies longitudinal force that stretches random paths favoring long trajectories. One can say that the condition $s\ge s^*$ selects a large-deviations regime where the path ensemble is conditioned on atypical, rare trajectories. At $s^* = \tfrac{1}{16}$ the system undergoes a thermodynamic phase transition. Beyond this threshold, the free energy density has a non-analytic behavior exhibiting a transition from a delocalized bulk diffusion to a localized boundary-pinned phase. These arguments recast the discussion of \eq{eq:15}, with an important difference: in \eq{eq:15} we considered  Brownian bridges sharing continuum spectrum without bound states, whereas here we deal with open but localized paths.

Further analysis of \eq{eq:25} deserves one more important comment. We shall consider the zero-momentum limit $\varkappa \to 0$. Setting $\varkappa = 0$ in \eq{eq:25} makes the probability density $P(y,\sigma)$ uniform in the $x$-direction. Physically, it means averaging out the horizontal transport and focusing entirely on how the random walk fluctuates along the radial "height" coordinate $y$. In the absence of the infrared screening length set by $\varkappa^{-1}$, the system becomes purely scale-invariant across the whole bulk domain $y \ge y_0$. The statistics of the ensemble of paths is governed exclusively by the competition between the entropic contribution from the volume of hyperbolic space and the microscopic UV cutoff at $y_0$. Introducing a small positive value $\varkappa^2 > 0$ generates a finite correlation length $\sim \varkappa^{-1}$ that competes with the intrinsic geometric scaling set by $\sigma$. Therefore, to simplify the consideration, we begin with the case $\varkappa = 0$.

Recall the spectral problem for the radial part of the 2D Schr\"odinger-like equation in a flat space $\mathbb{R}^2=\{(r,\varphi)\}$ in a conformally invariant attractive potential $-\lambda r^{-2}$ ($\lambda\ge 0$) with mixed (Robin) condition at the boundary of the disc of radius $R$: 
\be
\begin{cases}
\disp -E P(r,E, \lambda) = \left(d^2_{rr} + r^{-1} d_r\right) P(r,E,\lambda) + \lambda r^{-2}P(r,E,\lambda)=0 \medskip \\ 
\disp d_r P(r,E,\lambda) \Big|_{r=R} = \frac{g}{R} P(r=R,E,\lambda)
\end{cases}
\label{eq:28}
\ee
where $E$ is the spectral parameter. 

Imposing the mixed boundary condition is equivalent to adding a singular $\delta$-function potential, $-\frac{g}{R}\delta(r-R)$, at $r=R$ to the bulk potential $\lambda r^{-2}$. The normalization by $R$ ensures that the coupling constant $g$ is dimensionless. This boundary-value problem has been extensively studied in connection with scale-invariant inverse-square potentials and cyclic renormalization-group flows. In the supercritical regime, the RG flow develops a limit cycle, leading to discrete scale invariance and a geometric sequence of bound-state scales -- see, for example, \cite{kaplan, gorsky-RG, akkermans, hyp-efimov1, hyp-efimov2}. This structure is closely related to the mechanism underlying Efimov physics \cite{BawinCoon2003, MuellerHo2004, Efimov1970}. The solution is particularly illuminating in the limit $E\to 0$ in \eq{eq:28}, since the point $E=0$ probes the ground-state boundary, corresponding to the threshold of the critical binding--unbinding transition.

Upon making the identifications $\left\{r \Leftrightarrow y, \lambda \Leftrightarrow \sigma \right\}$, Eq. \eq{eq:25} at $\varkappa=0$ coincides with Eq. \eq{eq:28} at $E=0$. This establishes the dictionary between the spectral problem in the hyperbolic domain $\mathbb{H}^2$ ($y\ge y_0$) and the Schr\"{o}dinger-like problem in flat space $\mathbb R^2$, formulated in radial coordinates outside a disc of radius $R$ (i.e. for $r\ge R$). The resulting inverse-square problem is closely related to the scale-anomalous quantum-mechanical problem discussed in \cite{Case1950, BawinCoon2003}.

The generic RG equations, to first order in $\eps$, where $\eps$ parametrizes an infinitesimal change of the $y$-coordinate, $y \to y(1+\eps)$ ($0<\eps\ll 1$), were derived explicitly in \cite{akkermans}. Rather than reproducing the derivation here, it is more illuminating to use the solution of \eq{eq:28} to reconstruct the $\beta$-function, which depends on scale only implicitly through $g$ and describes the flow of the boundary coupling $g$ under the change of the spatial scale, $\ell = -\ln \tfrac{y}{R}$
\be
\beta(g) = \frac{d g}{d\ell}
\ee
Setting $E=0$ in Eq. \eq{eq:28} reduces it to a simple scale-invariant equation. Under the identifications defined above, we have:
\be
y^2 P''(y) + y P'(y) + \sigma P(y) = 0
\ee
For $y > 0$ the general real solution is:
\be
P(y) = C \sin\left(\sqrt{\sigma} \ln \frac{y}{R} + \theta\right); \quad (\sigma>0)
\ee
where $\theta$ is a phase shift. To define how coupling $g(y)$ changes across scales, consider the local scale-dependent logarithmic derivative:
\be
g(y) \equiv y \frac{P'(y)}{P(y)}
\ee
After a set of simple transformations we arrive at the equation for the $\beta$--function:
\be
\beta(g) =\frac{dg(\ell)}{d\ell}\equiv -\frac{dg(y)}{d\ln(y/R)} = (g^2 + \sigma) = (g - g_+)(g - g_-), \quad g_{\pm} = \pm i\sqrt{\sigma}
\label{eq:beta}
\ee
The solution of \eq{eq:beta} reads
\be
g(y) = \sqrt{\sigma} \cot\left(\sqrt{\sigma} \ln \frac{y}{R} + \theta\right)
\label{eq:29}
\ee 
where $\theta = \arctan\left(\frac{\sqrt{\sigma}}{g(R)}\right)\; \pmod{\pi}$. Equation \eq{eq:beta} implies the following regimes: 
\begin{itemize}
\item[(i)] For $\sigma<0$ the fixed points $g_{\pm}$ are real and distinct; 
\item[(ii)] For $\sigma=0$ the fixed points $g_{\pm}$ merge at $g_{+}=g_{-}=g_{c}=0$; 
\item[(iii)] For $\sigma>0$ the fixed points $g_{\pm}$ are complex conjugates and the flow of $g(y)$ is log-periodic, signaling the emergence of the discrete scale invariance \cite{BawinCoon2003, MuellerHo2004}.
\end{itemize}

Case (iii), with $\sigma>0$, is the most interesting for our purposes. Since the function $\cot(...)$ is periodic with period $\pi$, the function $g(y)$ returns to its initial value as $y$ returns to its initial value. Thus, the period of $g(y)$ is determined by the periodicity of the $\cot$--function:
\be
y(k)=y_0 \exp\left(\frac{\pi k}{\sqrt{\sigma}} + \theta\right), \quad k=1,2,3,...
\label{eq:30}
\ee
This discrete set exhibits the geometric scaling characteristic of discrete scale invariance and RG limit cycles \cite{BawinCoon2003, MuellerHo2004}:
\be
\Lambda = \Lambda_{UV}e^{\pi/\sqrt{\sigma}}
\label{eq:31}
\ee
where $\Lambda_{UV}=y_0 \exp({\theta})$ is the ultraviolet cutoff. Equation \eq{eq:31} exhibits the essential singularity typical for the Berezinskii-Kosterlitz-Thouless--type scaling \cite{Kosterlitz1974, Jose1977}: the correlation length $\Lambda$ diverges as $\sigma\to 0$. Note that the naive scaling consideration \eq{eq:17} based on the estimation of characteristic decay of the distribution function in $\mathbb{H}^2$ is consistent with the RG approach. The associated energy gap scale, $\Delta$, can be estimated as
\be
\Delta \sim \Lambda^{-2}\sim e^{-2\pi/\sqrt{\sigma}}
\label{eq:32}
\ee
In the next Section, we provide arguments supporting an interpretation of $\Delta$ as a Lifshitz-tail--like spectral scale, arising from rare fluctuations in a system without external quenched disorder as discussed in \cite{gorsky-nech-val} -- compare to \cite{Klopp2002, PasturFigotin1992} for the conventional Lifshitz-tail setting. 

Before discussing the BKT transition/Lifshitz tail from the perspective of the large-deviation principle, let us briefly comment on the connection between the RG equation \eq{eq:beta} and the RG equations of the two-dimensional XY model. In the 2D XY model near the vortex binding--unbinding transition, the Kosterlitz--Thouless renormalization-group equations for the temperature-like variable, $K \propto (T - T_{\text{BKT}})$, and vortex chemical potential, $\mu$, take the form \cite{Kosterlitz1974, Jose1977}:
\be
\begin{cases}
\dfrac{d K}{d\ln r} = -\mu^2, \medskip \\
\dfrac{d \mu}{d\ln r} = -K \mu
\end{cases}
\label{eq:xy}
\ee
Along the invariant surface $\mu^2 - K^2 = \sigma = \mathrm{const}$, the identification $g = K$ reduces Eqs. \eq{eq:xy} to a single effective RG equation:
\be
\frac{d g}{d\ln r} = -(g^2 + \sigma)
\label{eq:xy2}
\ee
identical (up to the sign \footnote{The difference in signs in Eqs. \eq{eq:beta} and \eq{eq:xy2} arises from the different definitions of the “RG time”, $\ell$: in our case, $\ell = -\ln y$, while in XY model $\ell$ is set as $\ell = \ln y$.}) to Eq. \eq{eq:beta}. The equation $\mu^2-K^2= \sigma= \mathrm{const}$ defines a two-dimensional hyperboloid embedded in the three-dimensional phase space, which locally represents a surface of constant negative curvature. The resulting hyperbolic geometry (the "saddle-like shape") provides a useful geometric representation of the RG flow and its separatrix structure. The correspondence with the BKT flow is therefore particularly transparent in this parametrization \cite{kaplan}.

Last but not least we ask what happens at $\varkappa>0$? To estimate the range of $\varkappa$ for which only the first terms of the BKT-like geometric sequence remain unscreened, we consider the radial equation at $E=0$ with a finite momentum $\varkappa$: $y^2 P''(y) + y P'(y) + (\sigma - \varkappa^2 y^2) P(y) = 0$. As we have seen, in the unscreened regime, for $y \ll \varkappa^{-1}$, the solutions exhibit log-periodic oscillations $y_k = \Lambda_{UV} \exp(\pi k / \sqrt{\sigma})$, whereas for $y \gg \varkappa^{-1}$ the mass term $-\varkappa^2 y^2$ forces exponential decay $P(y) \propto K_{i\sqrt{\sigma}}(\varkappa y) \sim e^{-\varkappa y}$, effectively providing the IR cutoff at $y \sim \varkappa^{-1}$. For only $k=1$ state (see \eq{eq:30}) to survive and to suppress all higher-order Efimov states ($k \ge 2$), the screening length must lie strictly between the first two levels, $y_1 \lesssim \varkappa^{-1}_c< y_2$, which gives $\Lambda_{UV} e^{\pi/\sqrt{\sigma}} \lesssim \varkappa^{-1} < \Lambda_{UV} e^{2\pi/\sqrt{\sigma}}$. Inverting these inequalities yields the precise parameter window $\Lambda_{UV}^{-1} e^{-2\pi/\sqrt{\sigma}} < \varkappa_c \lesssim \Lambda_{UV}^{-1} e^{-\pi/\sqrt{\sigma}}$, which identifies the critical upper threshold $\varkappa_c \sim \Lambda_{UV}^{-1} e^{-\pi/\sqrt{\sigma}}$ at which all higher bound states are screened, but the single-scale BKT asymptotic is preserved.

\section{Face 2: BKT--Lifshitz tail correspondence: deterministic large--deviations for spectral problem in $\mathbb{H}^2$}
\label{sect:03}

The presence of a "Lifshitz tail" (LT) in the eigenvalue density $\rho(E)$ near the spectral edge, $E_c$, of a $D$-dimensional disordered system is a well-known phenomenon in condensed matter and statistical physics \cite{lifshitz, pastur, luck}. Here we emphasize that a Lifshitz tail can also appear in diffusive systems without external disorder, arising from the interplay between discreteness, specific boundary conditions, and hyperbolic geometry, which together give rise to rare-event statistics \cite{gorsky-nech-val}. In the conventional Lifshitz-tail problem for a random Schr\"{o}dinger operator $H=-\nabla^{2}+V_{\mathrm{rand}}(r)$, the non-analytic density of states near the band edge $E_{c}$ originates from rare configurations of the random potential. Statistical weights of corresponding configurations are governed by the instanton (saddle-point) contribution to the associated action, $S_{\mathrm{opt}}$, leading to the following singular form of $\rho(E)$ in a D-dimensional space, which characterizes the Anderson localization:
\be
\rho(E)\sim e^{-S_{\mathrm{opt}}(E)}, \qquad S_{\mathrm{opt}}(E)\sim |E-E_{c}|^{-D/2}
\label{eq:38}
\ee
The precise form of the tail \eq{eq:38} is highly sensitive to the nature of the disorder. For example, a Poisson distribution gives rise to the Lifshitz tail in the form of \eq{eq:38}, whereas a Gaussian distribution of disorder, being smoother, does not produce exactly such a tail in $\rho(E)$ near the band edge.

Having in mind the 1D random trapping problem of Balagurov--Vaks \cite{balagurov}, and Donsker--Varadhan  \cite{donsker}, where the Lifshitz tail emerges due to the presence of external random distribution of defects, one may ask how the Lifshitz tail--like behavior manifests itself in the holographic disorder-free theory in $\mathbb{H}^2$. In \cite{gorsky-nech-val, grosb-nech-pol} it was conjectured that, qualitatively, the spatial disorder in the distribution of defects gets mapped onto a random distribution of “renewal times” of the diffusion near the boundary of a convex impermeable domain when diffusion time $t$ grows linearly with the domain size -- see Section \ref{sect:04} for detailed discussion of this question. 

Let us return to RG equation \eq{eq:beta} for the coupling $g(\ell)$, where $\ell = -\ln y$. This equation can be viewed as the deterministic dynamics of a particle moving in an effective potential $V(g)$:
\be
\frac{dg(\ell)}{d\ell} = \frac{dV}{dg}, \qquad V(g)=\tfrac{1}{3} g^{3} + \sigma\, g
\label{eq:33}
\ee
We can consider \eq{eq:33} as the optical/geometric formulation of the large-deviation principle for the RG flow. We reinterpret the logarithmic scale $\ell$ as an intrinsic geodesic length, $\tau$, along some one-dimensional manifold, $\mathcal{M}$, parameterized by the coupling constant $g$. Since $d\ell = \frac{dg}{\beta(g)}$, the natural element $d\tau^2$ induced by the flow dynamics could be written as:
\be 
d\tau^2 = \frac{dg^2}{\beta^2(g)} = \frac{dg^2}{(g^2 + \sigma)^2}
\ee
This description rhymes with the variational Fermat (eikonal) principle in geometrical optics \cite{eikonal} or to the WKB approximation in quantum mechanics. In this representation the RG "time" $\ell$ plays the role of an optical path length, while the inverse RG velocity $|dg/d\ell|^{-1}$ acts as an effective refraction index, $n(g) = \beta^{-1}(g)$. Such an interpretation is conceptually similar to the geometric optics description of buckled tissues discussed in \cite{pol-eikonal}. 

In the Riemannian geometry, the physical distance $d\tau = d\ell$ measures the "residence time" the moving particle spends around the fixed point. When $\sigma \to 0^+$, the point $g=0$ becomes a "bottleneck" for propagating particles. Near the BKT merging point ($g_+=g_-$) the flow essentially slows down. The RG trajectory must therefore traverse a region of a huge effective refractive index, and the accumulated "optical length" diverges as this region grows. This optical length plays the role of the \emph{eikonal action}. According to Fermat's principle in geometrical optics, or, similarly, to the eikonal equation 
\be
\left|\nabla S_{\mathrm{inst}}\right| = n(g),
\ee 
the optical path length, $S_{\mathrm{inst}}$, along the classical ray $\gamma$ connecting $g_{\rm initial} = -\infty$ to $g_{\rm final} = +\infty$ minimizes the functional:
\be
S_{\mathrm{inst}} = \int_{\gamma} n(g) ds = \int_{-\infty}^{\infty} \frac{dg}{\big|dg(\ell)/d\ell\big|} = \int_{-\infty}^{\infty} \frac{dg}{\beta(g)} = \int_{-\infty}^{\infty} \frac{dg}{g^2 + \sigma}
\ee
Evaluating this integral yields the eikonal large-deviation action. The "time" required for a trajectory to pass the nearly flat region of $V(g)$ in the vicinity of $\sigma=0$ is then:
\be
S_{\mathrm{inst}} = \left.\frac{1}{\sqrt{\sigma}} \arctan\left(\frac{g}{\sqrt{\sigma}}\right) \right|_{-\infty}^{+\infty} = \frac{\pi}{\sqrt{\sigma}}
\label{eq:instant1}
\ee
and the non-perturbative transition probability, $\mathcal{P}(\sigma)$ associated with the passage through this geometry-induced bottleneck can be estimated as:
\be
\mathcal{P}_{\mathrm{inst}}(\sigma) \sim e^{- S_{\rm inst}} = e^{-\frac{\pi}{\sqrt{\sigma}}}
\label{eq:instant2}
\ee
The essential singularity $e^{-\pi/\sqrt{\sigma}}$ represents a disorder-free, geometry-induced Lifshitz-like (LT) tail. The same structure appears in the WKB treatment of tunneling through a slowly varying potential barrier \cite{book-wkb}, where the transmission amplitude is governed by the exponential of the eikonal integral over the classically forbidden region. In the BKT problem the integral of the inverse RG velocity plays an identical role: it is the "action" of a deterministic instanton crossing the nearly flat region of $V(u)$. The singularity, which has a meaning of the lowest energy scale gap, can be estimated as in \eq{eq:32}: 
\be
\Delta \sim e^{-2 S_{\rm inst}}  \sim e^{-\frac{2\pi}{\sqrt{\sigma}}}
\label{eq:37}
\ee

The exponential suppression of the gap in \eq{eq:instant2} can be interpreted in terms of emergence of large--deviation regions in the RG potential landscape providing the cost of rare trajectories that pass across the flattened region of $V(g)$ to reach the strong--coupling regime. 

Comparing \eq{eq:instant2} and \eq{eq:37}, we suggest a connection between "WKB--BKT" and "Large deviation--LT" concepts, where LT emerges in a stochastic but disorder--free system driven into the large-deviation regime and the external disorder is replaced by a deterministic field that, in our particular example, is induced by the convex shape of the boundary. 

This derivation should be compared with the standard optimal fluctuation for the random cage model in the Balagurov and Vaks setting \cite{balagurov}. We split the free energy into two parts, $F=F_{\rm conf}+F_{\rm ent}$. The part $F_{\rm conf}$ accounts for $t$-step path's confinement $t$ in a spontaneously created cage ("void") of a dimensionless volume $(R/a)^D$ in a $D$-dimensional space and can be estimated as $F_{\rm conf}(R)/(k_BT) \sim t E_{\rm min}(R)$, where $E_{\rm min}(R)$ is the minimal eigenvalue (i.e. "energy") of the Laplace operator in the domain $\mathbb{R}^D$, which in typical geometries is $E_{min}(R)\sim 1/R^2$. So, one has $F_{\rm conf}(R) \sim \frac{t}{R^2}$. The part $F_{\rm ent}(E)$ for the Poissonian disorder can be estimated as $F_{\rm ent}/(k_BT) \sim -\ln e^{-b R^D} \sim b R^D$ where  $e^{-b R^D}$ is the probability of a spontaneous creation of an empty void of typical dimensionless size $R$ in a $D$-dimensional space and $b$ is a coefficient accounting for the density. The expression of $F_{\rm ent}$ is the manifestation of the statistics of rare events. Finding a minimum of $F(R)=F_{\rm conf}(R)+F_{\rm ent}(R)$ over $R$  we get size of an "optimal" cage 
\be
R_{\rm opt}(r) \sim \left.\left(\frac{2t}{D b}\right)^{1/(D+2)}\right|_{D=1} = \left(\frac{2t}{b}\right)^{1/3} 
\label{eq:opt01}
\ee
Tracing this expression back to the survival probability, $P(t)\sim e^{-F(R_{\rm opt})/(k_BT)}$, we get in one dimension ($D=1$):
\be
P(t) \sim e^{-3 b R_{\rm opt}(t)} \sim e^{-\gamma\, b^{2/3}t^{1/3}}  
\label{eq:lt}
\ee
where $\gamma=\tfrac{3}{2^{2/3}}$ is the non-universal model-dependent constant. In Eq. \eq{eq:sc02}  we reproduce the stretched exponential scaling \eq{eq:lt} by pushing the ensemble of paths that evade the disc to the large-deviation regime via stretching without any external randomness.

By the Legendre transform, implemented as the inverse Laplace (Mellin) transform, we pass from the microcanonical ensemble of trajectories in $t$ to the canonical ensemble where $t$ is controlled by the chemical potential ("energy"), $E$, attributed to each step of a path. This leads us to the grand canonical form of the survival probability \cite{nieuwen}, which gives the desired spectral density, $\rho(E)$. So, with exponential accuracy we get:
\be
\rho(E) = \frac{1}{2\pi i}\oint P(t)\, e^{E t}\, dt \sim  e^{-\gamma^{3/2}\, b/\sqrt{E}}
\label{eq:sc04}
\ee
As one sees, $\rho(E)$ has the form of the Lifshitz tail \eq{eq:38} which is a signature of 1D Anderson localization, where the constant $b$ controls the disorder strength.

Returning to \eq{eq:37}, we can conclude that the exponentially small scale $\Delta$, governed by the action $S_{\mathrm{inst}}\propto 1/\sqrt{\sigma}$, is analogous to the 1D optimal--fluctuation action $S_{\mathrm{opt}}(E)$. The RG potential $V(g)$ acts as a "deterministic large--deviation landscape" for the RG flow responsible for the LT. The BKT singularity and the LT thus share a common mathematical origin: both are governed by the instanton--type configuration in a nearly flat potential. 

In summary, from the RG perspective, the “deterministic rare event” can be viewed as an anomalously long excursion of the RG flow near the marginal fixed point, whereas in a truly disordered system it corresponds to a rare spatial region capable of supporting a localized state. The concept of a “deterministic large-deviation landscape” was introduced previously in works devoted to the similarity between the Lifshitz tails and the survival probability of a stretched 2D random walk above an impermeable disc \cite{gorsky-nech-val, grosb-nech-pol}. In the next section, we discuss existing and new results for spectral problem in the Poincar\'{e} bounded domain.

\section{Face 3: KPZ-like scaling of stretched paths above the disc in $\mathbb{R}^2$ and WKB asymptotics in bounded Poincar\'{e} domain $\mathbb{H}^2$}
\label{sect:04}

In this Section we discuss a boundary problem for stretched diffusion above an impermeable disc in the Euclidean plane. The latter formulation has a straightforward connection with the one-dimensional KPZ scaling behavior. Thus, relying on the similarity of diffusion in the hyperbolic plane $\mathbb{H}^2 = \{(x, y) | y>y_0\}$ and in the Euclidean plane in polar coordinates $(r, \varphi)$, we provide arguments for the emergence of KPZ-like scaling for stretched diffusion along $x$-coordinate in $\mathbb{H}^2$. 

The motivation to study diffusion above a convex void was inspired by H. Spohn and P. Ferrari in \cite{ferrari}, where the authors analyzed the statistics of one-dimensional directed random walks constrained to stay above a quadratic algebraic curve. It is known that the fluctuations of the top line in a bundle of one-dimensional directed “vicious walks” (equivalently, the world lines of free fermions in 1D with fixed endpoints) are governed by the Tracy--Widom distribution \cite{tracy}. Following the approach of \cite{ferrari}, one can define the average position of the top line and study its fluctuations. In this picture, all the vicious walks lying below the top line act as a “mean field” representing the bulk, which pushes the top line toward an equilibrium position. Fluctuations around this equilibrium differ from those of a free random walk in the absence of the bulk. Replacing the bulk’s effect with a circular boundary leads to a model in which a one-dimensional directed random walk is constrained to stay above the semicircle, whose interior is inaccessible to the path. In \cite{ferrari}, the authors demonstrated that this system belongs to the KPZ universality class.

There is a debatable question whether the observed "1/3--scaling" is a signature of KPZ, or just an occasional coincidence, and whether one could provide arguments beyond the Spohn--Ferrari model. Our point of view is that, although the observed scaling for fluctuations of paths does not share the full Tracy--Widom distribution, it nevertheless inherits the main ingredients of actual KPZ statistics, being related to the particular case of the Painleve II equation that lies at the heart of the Tracy-Widom analysis. Since this question is actually important but lies outside the main line of the current work, we leave this discussion for the Appendix \ref{sect:app1}. Throughout the text, we use the neutral formulation “KPZ-like scaling”.

\subsection{Stretched diffusion evading the disc}
\label{sect:04a}

\begin{figure}[hb]
    \centering
    \includegraphics[width=0.5\linewidth]{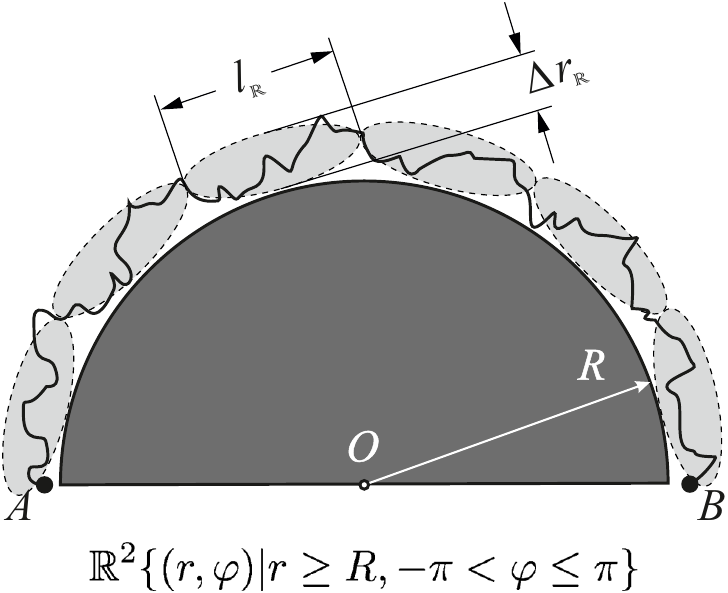}
    \caption{Brownian bridge above the semicircle of radius $R$ in $\mathbb{R}^2$ stretched along boundary; $\Delta r_{\mathbb{R}}$ and $l_{\mathbb{R}}$ are the typical "span" and the "correlation lengths". The path is split in "elliptic blobs" which schematically designate uncorrelated parts.}
    \label{fig:02}
\end{figure}

Here, we analyze the statistics of diffusive trajectories constrained to stay above a semicircle under various boundary conditions where the statistics of rare events arise naturally. Schematically the considered model is depicted in \fig{fig:02}.

Specifically, we consider the Cauchy problem for diffusion in polar coordinates, subject to Dirichlet, or mixed (Robin), boundary conditions at $r=R$:
\be
\begin{cases}
\disp \partial_t P(r,\varphi,t)= \tfrac{1}{4}\left(r^{-1}\,\partial_r\bigl(r\,\partial_r P\bigr) + r^{-2}\,\partial_{\varphi}^2 P\right) \medskip \\ 
\disp  P(R,\varphi,t)=0 \mbox{ or } \partial_r P(R,\varphi,t)+\kappa\,P(R,\varphi,t)=0 \medskip \\
\disp P(r,0,t)=P(r,\pi,t)=0,\quad \lim_{r\to\infty}P(r,\varphi,t)=0.
\end{cases}
\label{eq:rwdrift}
\ee
where one assumes zero conditions on the rays $\varphi=0$ and $\varphi=\pi$. The initial condition fixes the starting point of the diffusion $(r_0, \varphi_0)$:
\be
P(r,\varphi,0)=r_0^{-1}\,\delta_{r-r_0}\,\delta_{\varphi-\varphi_0},
\qquad r_0=R^+,\quad \varphi_0=\pi.
\label{eq:rwdrift3}
\ee

Following the setting considered in \cite{nech-pol-val}, the probability of a Brownian bridge to stay above the top of the semicircle is defined by the probability of joining two independent propagators at $\varphi=\tfrac{\pi}{2}$: (i) the "forward" propagator  $P_c^{\to}(r,\tfrac{\pi}{2},t_0\mid r_0,\varphi_0)$, and (ii) the "backward" one $P_c^{\leftarrow}(r,\tfrac{\pi}{2},t_0\mid r_1,\varphi_1)$, where $r_0= R^+, \varphi_0=\pi^-$ and $r_1 = R^+, \varphi_1 =0^+$; $P_c^{\to}(...)$ and $P_c^{\leftarrow}(...)$ are equal due to the symmetry of our problem. We also impose the stretching condition for the time of the total Brownian excursion (the total length of the path): $t=c R$. Thus the total probability of finding a path at the point $(r, \tfrac{\pi}{2})$ is:
\be
\Theta(r;R,c,t)= P^2_c\big(r,\tfrac{\pi}{2},\tfrac{t}{2}\mid r_0,\varphi_0\big),
\label{eq:rwdrift4}
\ee
After normalization, we get the conditioned partition function $\Omega(r;R,c,t)$
\be
\Omega(r;R,c,t)=\frac{\Theta(r;R,c,t)}{\displaystyle\int_0^\infty \Theta(s;R,c,t)\,ds}.
\label{eq:rwdrift5}
\ee
using which we compute the average span of the random walk above the disc of radius $R$ in the Euclidean plane $\mathbb{R}^2$: 
\be
\Delta r_{\mathbb{R}}(R) \equiv \sqrt{\langle r^2_{\mathbb{R}}(R)\rangle} = \sqrt{\int_{R}^{\infty} \Omega(r;R,c,t) r^2 dr - \left(\int_{R}^{\infty} \Omega(r;R,c,t) r dr\right)^2}
\label{eq:rwdrift6}
\ee

\begin{figure}[ht]
    \centering
    \includegraphics[width=0.95\linewidth]{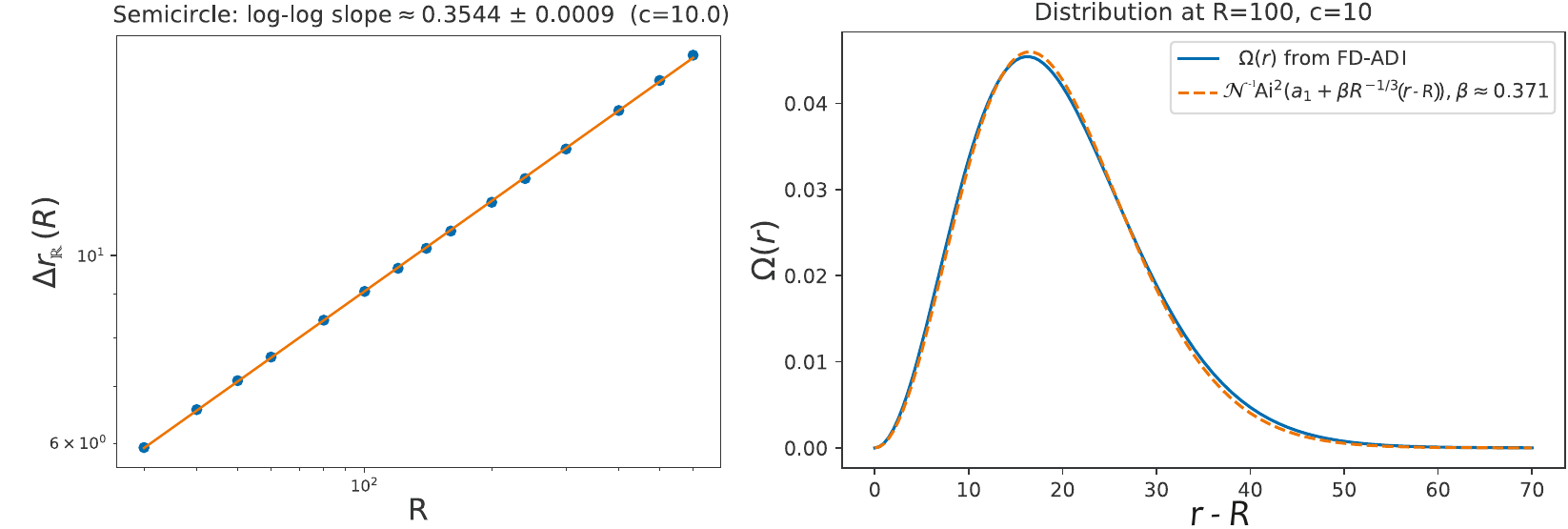}
    \caption{Left: $\Delta r_{\mathbb{R}}(R)$ as a function of $R$ for stretched paths $t = c R$ in log--log scale for Dirichlet condition. The uncertainty for the fitted exponent is the standard error of the linear regression. No error bars are shown for the individual points, since each point is computed deterministically from the numerical solution of the PDE; Right: Comparison of $\Omega(r)$ with $\mathcal{N}^{-1} \mathrm{Ai}^2(\beta R^{-1/3}(r-R)+a_1)$ for $R=100$, $c=10$ and Dirichlet boundary conditions; $a_1\approx -2.338...$, $\beta\approx 0.371$.}
    \label{fig:03}
\end{figure}

\begin{figure}[ht]
    \centering
    \includegraphics[width=0.95\linewidth]{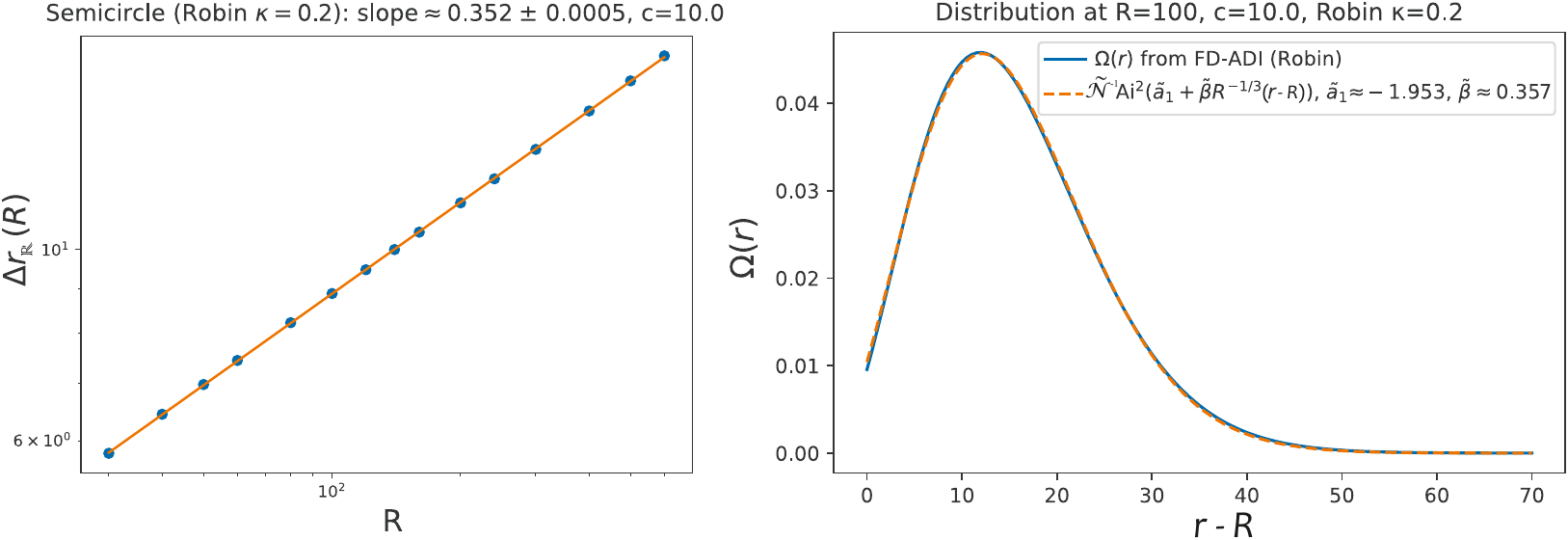}
    \caption{Left: $\Delta r_{\mathbb{R}}(R)$ as a function of $R$ for stretched paths $t = c R$ in log--log scale for Robin condition. The uncertainty for the fitted exponent is the standard error of the linear regression. No error bars are shown for individual points, since each point is computed deterministically from the numerical solution of the PDE; Right: Comparison of $\Omega(r)$ with $\tilde{\mathcal{N}}^{-1} \mathrm{Ai}^2(\tilde{\beta} R^{-1/3}(r-R)+\tilde{a}_1)$ for $R=100$, $c=10$ and mixed boundary conditions with $\kappa=0.2$; $\tilde{a}_1\approx -1.953$, $\tilde{\beta}\approx 0.357$.}
    \label{fig:04} 
\end{figure}

We solve the Cauchy problem \eq{eq:rwdrift}--\eq{eq:rwdrift3} numerically using a finite-difference scheme for the propagator $P_{\rm c}\big(r,\tfrac{\pi}{2},t/2\,\big|\,r_0,\varphi_0\big)$. This PDE is treated using Peaceman–Rachford algorithm (alternating--direction implicit method) \cite{peacemanrachford}: in the first half--step, the scheme is implicit in $r$ and explicit in $\varphi$; in the second half--step, the scheme is implicit in $\varphi$ and explicit in $r$. The initial condition is a $\delta$--spike in the first radial cell near $r=R$ at $\varphi=\pi$. For the total time $t$, we propagate to $\tfrac{t}{2}$. Following this procedure, one can straightforwardly compute the function $\Theta(r;R,c,t)$ using \eq{eq:rwdrift4} and then evaluate $\Delta r_{\mathbb{R}}(R)$ for two different boundary conditions: (i) Dirichlet, $P(R)=0$, and (ii) mixed (Robin), $\partial_r P(r)|_{r=R}+\kappa P(R)=0$. The corresponding results for $\Delta r_{\mathbb{R}}(R)$ and the radial conditional distribution of the Brownian bridge middle point, $\Omega(r;R,c,t)$, are shown in \fig{fig:03} for the Dirichlet boundary condition (i), and in \fig{fig:04} for the mixed boundary condition (ii).

Let us briefly describe the implementation of the realized numerical algorithm. The computations are performed on the polar grid $\rho \in [R,R+r_{\max}], \varphi\in[0,\pi]$ with $R=30 \div 600$ with the stretching parameter $c=10$, radial mesh size $\Delta r=0.5$, and time step $\Delta t=2.5$. Since the Brownian bridge is constructed by joining two identical half-propagators at
$\varphi=\pi/2$, each finite-difference solve is performed only up to $t/2=cR/2$. With $c=10$ and $\Delta t=2.5$, this gives $n_t=\frac{t_{\rm half}}{\Delta t} =\frac{cR}{2\Delta t} =2R$. For the Dirichlet condition we have $r_{\max}\simeq 10\sqrt{R}$ and $N_\varphi \simeq 128\sqrt{R/100}$, while for the Robin condition we have $r_{\max}\simeq 12\sqrt{R}$ and $N_\varphi \simeq 1.3\times 128\sqrt{R/100}$. The scaling $r_{\max}\sim \sqrt{R}$ keeps the outer boundary at a fixed
number of diffusive lengths, since $t_{\rm half}\sim R$. The angular
resolution was increased as $N_\varphi \sim \sqrt{R}$ in order to keep the
discretization of the angular part stable over the whole range of $R$. The larger cutoff and angular resolution in the mixed-boundary case were used to resolve the Robin boundary layer more accurately and to suppress residual sensitivity to the outer boundary. The singular initial condition is regularized by a Gaussian function localized in the first interior cell near the boundary, with widths $\sigma_r=\Delta r$ and $\sigma_\varphi=\Delta\varphi$, and is normalized with respect to the measure in polar coordinates. 

The mixed boundary conditions are realized via the finite-difference scheme as $\tfrac{P_1-P_0}{\Delta r}=\tfrac{\kappa}{D}P_0$ and $P_0=\tfrac{P_1}{1+\kappa\Delta r/D}$. The sensitivity of the fitted exponent to the fitting interval was checked by repeating the least-squares fit of $\log \Delta r_{\mathbb{R}} = {\rm const}+\gamma \log R$. For the full interval $30\le R\le 600$, we obtain $\gamma_{\rm D}=0.3544, 95\%~{\rm CI}=[0.3524,0.3564]$ for the Dirichlet condition, and $\gamma_{\rm R}=0.3525, 95\%~{\rm CI}=[0.3514,0.3536]$ for the Robin condition. Variation of the lower fitting cutoff gives us: $\gamma_{\rm Dirichlet}=0.354\ldots 0.364$, and $\gamma_{\rm Robin}=0.352\ldots 0.357$. These results are summarized in the Table \ref{tab:fit-sensitivity}.

\begin{table}[h]
\centering
\begin{tabular}{c|c|c|c} 
\hline\hline
Boundary condition & full-fit $\gamma$ & 95\% CI & cutoff sensitivity \\ \hline
Dirichlet & $0.3544$ & $[0.3524,0.3564]$ & $0.354\ldots0.364$ \\ \hline
Mixed (Robin), $\kappa=0.2$ & $0.3525$ & $[0.3514,0.3536]$ & $0.352\ldots0.357$ \\
\hline\hline
\end{tabular}
\caption{Sensitivity of the fitted exponent in $\Delta r_{\mathbb{R}}(R)\sim R^\gamma$ to the fitting interval.}
\label{tab:fit-sensitivity}
\end{table}

We also checked the stability of the numerical result at $R=100$ by varying
the main discretization parameters one at a time. The radial cutoff was changed
from the value mentioned above to both a smaller and a larger value,
$r_{\max}\simeq 8\sqrt{R}$ and $r_{\max}\simeq 12\sqrt{R}$ in the
Dirichlet case, and $r_{\max}\simeq 10\sqrt{R}$ and
$r_{\max}\simeq 14\sqrt{R}$ in the mixed-boundary case. The time step was
changed from $\Delta t=2.5$ to $\Delta t=5.0$ and $\Delta t=1.25$, and
the radial mesh size from $\Delta r=0.5$ to $\Delta r=1.0$ and
$\Delta r=0.25$. The angular resolution was also varied: for the Dirichlet
case we changed $N_\varphi$ from its main value $N_\varphi=128$ to
$N_\varphi=96$ and $N_\varphi=192$, while for the mixed-boundary case we
changed it from $N_\varphi=168$ to $N_\varphi=128$ and $N_\varphi=206$.
In all these tests the computed value of $\Delta r_{\mathbb{R}}$ changed only
weakly; the largest variation came from the angular resolution and remained at
the level of about $0.5\%\div 2\%$. Thus the observed scaling is stable under reasonable changes
of the numerical parameters.

In both cases, (i) and (ii), the distribution function and the average span of radial fluctuations of stretched paths $t=c R$ for $c=10$ have the following asymptotic expressions:
\be
\Omega(r;R,c,t) =\frac{1}{\mathcal{N}}\, \mathrm{Ai}^2\left(\frac{\beta (r-R)}{R^{1/3}} + a_{\rm bound}\right), \qquad \Delta r_{\mathbb{R}}(R) \sim c^{2/3} R^{1/3} \sim c^{1/3} t^{1/3}
\label{eq:sc03}
\ee
where the normalization $\mathcal{N}$, $a_{\rm bound}$ and $\beta$ are boundary-dependent numerical constants. In particular, the boundary-dependent shift, $a_{\rm bound}$ takes the values $a_{\rm bound} = \{a_1, \tilde{a}_1\}$: $a_1\approx -2.338...$ is the first zero of the Airy function for the Dirichlet condition \cite{ferrari}, and $\tilde{a}_1\approx -1.953$ is the shifted value of $a_1$ for the Robin condition. In Appendix \ref{sect:app2}, we briefly reproduce the analytical derivation of the distribution function $\Omega(r;R,c,t)$ in \eq{eq:sc03} for the Dirichlet condition, following the approach proposed in \cite{fedotov-nech}. It seems instructive to recall the main steps of the derivation here, as they are based on the large-deviation principle and are consistent with the WKB-type framework discussed at length in Section \ref{sect:04b}.

The dependence $\Delta r_{\mathbb{R}}(R) \sim c^{2/3} R^{1/3}$ in \eq{eq:sc03} is  consistent with the scaling arguments of \cite{nech-pol-val,grosb-nech-pol, fedotov-nech}. Here we briefly recall them. As shown in \fig{fig:02}a, split the entire stretched path of length $N$ running from $A$ to $B$ above the semicircle into a train of independent "blobs" whose longitudinal size, $l_{\mathbb{R}}(R)$, is determined by the condition that, for a given radius $R$, the fragment of the convex boundary of the order of the blob's length, $l_{\mathbb{R}}(R)$, can be regarded as a flat segment. A stretched path follows a straight line as much as possible and becomes curved only when the curvature cannot be avoided. A random path that has to travel a horizontal distance $l_{\mathbb{R}}$ is localized within a strip of typical vertical span $\Delta r_{\mathbb{R}}= \sqrt{l_{\mathbb{R}}}$. If the path is forced to travel a distance $l_{\mathbb{R}}$ along a curved arc, and the arc fits within this strip, the curvature of the arc can be neglected. Therefore, the arc length whose curvature can be ignored is defined by the equation $l_{\mathbb{R}}^2/(c^2 R) \sim \sqrt{l_{\mathbb{R}}}$. Solving this equation, we get $l_{\mathbb{R}}\sim c^{4/3} R^{2/3}$. At shorter distances, the stretched path can be regarded as an unconstrained random walk. The span in the vertical direction is $\Delta r_{\mathbb{R}} \sim \sqrt{l_{\mathbb{R}}} \sim c^{2/3} R^{1/3}$. Each blob of length $l_{\mathbb{R}}$ defines a typical "renewal distance". Beyond this renewal distance, the arc itself deviates considerably from a straight segment. These estimates allow us to guess the scaling dependence of the free energy, $F(t)$ of ensemble of random paths of length $t$ stretched above the disc. Recall that stretching means that "lengths" $t$ grows linearly with the disc's radius, $R$, i.e. $t=cR$, where $c$ is stretching constant. The free energy can be estimated as the number of "elliptic blobs", $m=t/l_{\mathbb{R}}$, which is fully consistent with the optimal fluctuation approach provided in \cite{fedotov-nech}:
\be
\frac{F(t)}{k_BT}\bigg|_{t = cR} \sim \frac{t}{l_{\mathbb{R}}} \sim \frac{R^{1/3}}{c^{1/3}}  \sim \frac{t^{1/3}}{c^{2/3}}
\label{eq:sc02}
\ee

In \cite{fedotov-nech} we used the optimal fluctuation approach for stretched paths, $t = cR$ staying above the impermeable disc. Balancing the confinement, $F_{\rm conf}$ and the entropic, $F_{\rm ent}$, contributions to the free energy, we argued that the radial first-return ("survival") probability $P(t)$ with the exponential accuracy behaves as 
\be
P(t) \sim e^{-\frac{F(t)}{k_BT}} \sim e^{-\mathrm{const}\, \frac{t^{1/3}}{c^{2/3}}}
\label{eq:sc01}
\ee 
The asymptotics \eq{eq:sc01} has appeared in the literature under various names, such as the "stretched exponent", "Griffiths singularity", "Balagurov–Vaks trapping exponent", etc. However, as noted in \cite{nieuwen}, in all these cases it represents nothing else than the inverse Laplace-transformed Lifshitz tail of one-dimensional disordered systems exhibiting Anderson localization. In the random trapping problem, such as the Balagurov–Vaks one, the stretched exponential behavior arises due to the spontaneous emergence of rare, trap-free regions within a \emph{disordered} medium. In contrast, the present system is entirely free of disorder, and the non-algebraic decay is instead attributed to large-deviation effects resulting from the interplay between curvature and stretching. 

Comparing \eq{eq:sc01} and \eq{eq:lt} suggests the connection $c\sim b^{-1}$, where $c$ represents the stretching and $b$ denotes the disorder strength. The physical argument behind this connection is that stronger path stretching effectively corresponds to a system with sparser disorder. Thus, the Lifshitz tail in the specific "disorder-free" system, driven into a large-deviation regime near curved boundary, can be viewed as the "grand canonical manifestation" of the stretched exponential behavior of the survival probability associated with KPZ-like scaling.

\subsection{Spectral problem in $\mathbb{H}^2$ via WKB approximation}
\label{sect:04b}

To connect Sections \ref{sect:02} and \ref{sect:03}, we study the spectral problem in $\mathbb{H}^2$ subject to a Dirichlet boundary condition at $y_0$. Our primary goal is to derive the spectrum of path fluctuations in the vicinity of $y_0$ using the WKB approximation. We return to the diffusion process in $\mathbb{H}^2$ given by \eq{eq:09}. By separating variables and applying the ansatz $Q(x,y,t)=e^{-st-i\varkappa x}Q(y)$ to \eq{eq:09}, we obtain the spectral problem for $Q(y)$ in $\mathbb{H}^2$:
\be
-4s Q(y)=y^{2}\left(d_{yy}^2-\varkappa^{2}\right)Q(y)
\label{eq:wkb01}
\ee
This equation is similar to \eq{eq:23}, except that here we consider a Dirichlet boundary condition rather than mixed boundary conditions in \eq{eq:23}.

Instead of deriving the exact solution of \eq{eq:wkb01} equipped with the Dirichlet boundary condition, we shall seek an approximate solution of \eq{eq:wkb01} using the WKB quasiclassical approach \cite{book-wkb, olver-1974}. The motivation for this "WKBfication" of the boundary problem is as follows: we aim to show that the subset of paths providing the dominant contribution to the quasiclassical solution of \eq{eq:wkb01} is the same large-deviation ensemble of "stretched" paths evading the impermeable disc in the radial $\mathbb{R}^2$ problem considered in Section \ref{sect:04a}.

To proceed with the WKB approach, rewrite \eq{eq:wkb01} as follows:
\be
Q''(y)=G(y)Q(y), \qquad G(y)=\varkappa^{2}-\frac{4s}{y^{2}}=\varkappa^{2}-\frac{\sigma}{y^{2}}, \qquad \sigma=4s
\label{eq:wkb02}
\ee
where $Q''(y) = \frac{d^2 Q(y)}{dy^2}$.

The turning point $y_{\rm turn}$ of \eq{eq:wkb02} is defined by the condition $G(y=y_{\rm turn})=0$ \cite{olver-1974}, which gives $y_{\rm turn}=\frac{\sqrt{\sigma}}{\varkappa}$. Near the turning point we expand $G(y)$ to first order:
\be
G(y)\approx G'(y_{\rm turn})(y-y_{\rm turn}), \qquad G'(y)=\frac{2\sigma}{y^{3}}
\label{eq:wkb03}
\ee
where $G'(y) = \frac{dG(y)}{dy}$. Hence we get
\be
G'(y_{\rm turn})=\frac{2\sigma}{y_{\rm turn}^{3}}=\frac{2\sigma}{(\sqrt{\sigma}/\varkappa)^{3}} =\frac{2\varkappa^{3}}{\sqrt{\sigma}}
\label{eq:wkb04}
\ee

Introducing the rescaled variable $\eta = \left(G'(y_{\rm turn})\right)^{1/3}(y-y_{\rm turn})$, we rewrite \eq{eq:wkb02} in the form of the Airy equation \cite{book-wkb, olver-1974}:
\be
Q''(\eta) - \eta Q(\eta) = 0
\label{eq:wkb05}
\ee
Near $y=y_{\rm turn}$, the solution (which also respects the behavior at infinity) is
\be
Q(\eta)\approx C_1 \mathrm{Ai}(\eta), \qquad  \eta = \big(G'(y_{\rm turn})\big)^{1/3}(y-y_{\rm turn})
\label{eq:wkb06}
\ee
where $C_1$ is the normalization constant and $\mathrm{Ai}(\dots)$ is the Airy function. Imposing the Dirichlet condition, we have
\be
Q(\eta_0)=0 \quad\Rightarrow\quad \mathrm{Ai}(\eta_0)=0, \quad \eta_0 = \left(G'(y_{\rm turn})\right)^{1/3}(y_0-y_{\rm turn})
\label{eq:wkb07}
\ee
Because all real zeros of the Airy function are strictly negative, we represent the first zero as $-|a_1|$, where $|a_1|\approx2.338107$. The boundary condition $Q(\eta_0)=0$ is thus respected when $\mathrm{Ai}(\eta_0=-|a_1|)=0$. The equation $\eta_0=-|a_1|$ gives
\be
\left(G'(y_{\rm turn})\right)^{1/3}(y_0-y_{\rm turn}) = -|a_1|
\label{eq:wkb08}
\ee
Summarizing the obtained expressions, we have:
\be
G'(y) = \frac{2\sigma}{y^{3}}, \quad y_{\rm turn} = \frac{\sqrt{\sigma}}{\varkappa}, \quad G'(y_{\rm turn}) = \frac{2\varkappa^{3}}{\sqrt{\sigma}}
\label{eq:wkb09}
\ee
Therefore
\be
\big(G'(y_{\rm turn})\big)^{1/3} = \left( \frac{2\varkappa^{3}}{\sqrt{\sigma}} \right)^{1/3} = 2^{1/3}\frac{\varkappa}{\sigma^{1/6}}
\label{eq:wkb10}
\ee
Substituting \eq{eq:wkb10} into the Airy boundary condition gives us
\be
2^{1/3}\frac{\varkappa}{\sigma^{1/6}} \left(y_0 - \frac{\sqrt{\sigma}}{\varkappa} \right) = -|a_1| \quad \Rightarrow \quad 2^{1/3} \varkappa y_0 - 2^{1/3} \sigma^{1/2} = -|a_1| \sigma^{1/6}
\label{eq:wkb11}
\ee
Introducing $u = \sigma^{1/6}$ and $\alpha= 2^{-1/3}|a_1| > 0$, the condition \eq{eq:wkb11} yields $2^{1/3}\varkappa y_0 - 2^{1/3}u^3 = -2^{1/3}\alpha u$. Dividing by $2^{1/3}$ and arranging terms, it becomes the cubic equation
\be
u^{3} - \alpha u - \varkappa y_0 = 0
\label{eq:wkb12}
\ee
Selecting the real root in Cardano's formula for $u^{3} + p u + q = 0$ with $p=-\alpha$ and $q=-\varkappa y_0$, we get:
\be
u =\left(\frac{\varkappa y_0}{2} + \Delta^{1/2}\right)^{1/3} + \left(\frac{\varkappa y_0}{2} - \Delta^{1/2}\right)^{1/3}, \quad \Delta = \left(\frac{\varkappa y_0}{2}\right)^{2} - \left(\frac{\alpha}{3}\right)^{3}, \quad \Delta >0
\label{eq:wkb13}
\ee
For the large--$y_0$ regime considered below, $\varkappa y_0\gg 1$, and hence $\Delta>0$, so this real-root representation is sufficient. The Airy argument near the turning point is
\be
\eta(y) = \big(G'(y_{\rm turn})\big)^{1/3}(y-y_{\rm turn}) = 2^{1/3}\frac{\varkappa}{u} \left(y - \frac{u^{3}}{\varkappa} \right)
\label{eq:wkb14}
\ee
therefore the solution of the spectral equation near the turning point in $\mathbb{H}^2$ is
\be
Q(y)\propto  \mathrm{Ai}\left(\frac{\varkappa 2^{1/3}}{u}\left(y - \frac{u^3}{\varkappa}\right) \right)
\label{eq:wkb15}
\ee
where $u$ is implicitly expressed through $y_0$ using \eq{eq:wkb13}.

For large $y_0$ when $\varkappa y_0\gg 1$, the systematic asymptotic expansion of $u$, is
\be
u = (\varkappa y_0)^{1/3} + \frac{\alpha}{3(\varkappa y_0)^{1/3}} + O\left((\varkappa y_0)^{-5/3}\right)
\label{eq:wkb16}
\ee
The width of the Airy region follows from $\Delta\eta\sim 1$, which gives
\be
\Delta y_{\mathbb H} \sim \big(G'(y_{\rm turn})\big)^{-1/3} = \frac{u}{2^{1/3}\varkappa}
\label{eq:wkb17}
\ee
Using the large--$y_0$ expansion of $u$ from \eq{eq:wkb16}, the width of the Airy region is
\be
\Delta y_{\mathbb H} = 2^{-1/3}\varkappa^{-2/3}y_0^{1/3} + \frac{|a_1|}{3\times 2^{2/3}}\varkappa^{-4/3}y_0^{-1/3}  + O\left(y_0^{-5/3}\right)
\label{eq:wkb18}
\ee
so in the leading order we recover
\be
\Delta y_{\mathbb H} \approx 2^{-1/3}\varkappa^{-2/3}y_0^{1/3}
\label{eq:wkb19}
\ee

The Airy-based solution appears naturally as the local analysis in the WKB approach \cite{book-wkb}. Away from the turning point $y_{\rm turn}$, where the effective function $G(y)=\varkappa^{2}-\sigma y^{-2}$ does not vanish, the solution is approximated by the standard WKB expression in the region $G(y)>0$:
\be
Q_{\mathrm{WKB}}(y)\sim G(y)^{-1/4}\exp \left(\pm\int^{y}\sqrt{G(\mu)}\, d\mu\right).
\ee
In the region $G(y)<0$, the corresponding WKB solution is oscillatory and can be written as
\be
Q_{\mathrm{WKB}}(y)\sim |G(y)|^{-1/4}\exp \left(\pm i\int^{y}\sqrt{|G(\mu)|}\, d\mu\right).
\ee
However, in a small neighborhood of $y_{\rm turn}$ the standard WKB approximation fails, and the function $G(y)$ should be expanded in the vicinity of $y_{\rm turn}$ up to the first linear term. Writing $G(y)\approx G'(y_{\rm turn})(y-y_{\rm turn})$ brings the diffusion equation into the Airy equation form. The resulting Airy solution provides the uniform interpolation between the oscillatory and exponential WKB branches. 

The Dirichlet boundary condition $Q(y_0)=0$ is imposed by requiring the Airy function to vanish at the boundary. For the lowest mode, this corresponds to matching the Airy argument to its first zero,
$\mathrm{Ai}(-|a_1|)=0$ -- see \cite{olver-1974}. Thus, the Airy-zero condition provides the turning-point form of the WKB eigenvalue condition for the lowest mode. Since $-|a_1|<0$, this condition places the boundary $y_0$ on the oscillatory side of the turning point, while the solution for $y>y_{\rm turn}$ belongs to the exponentially decaying branch.

It is instructive to compare the numerical solution of stretched diffusion above the impenetrable disc given by the Airy distribution \eq{eq:sc03} in $\mathbb{R}^2$ with the WKB-type approximate solution of diffusion in $\mathbb{H}^2$ given by \eq{eq:wkb15}. These distributions provide similar results up to a redefinition of the variables $R \Leftrightarrow y_0$, as discussed in Section \ref{sect:02}. The distribution in \eq{eq:sc03} is squared in the Airy function because it corresponds to the inner part of the Brownian bridge and is "glued" from two independent parts. The physical assumptions behind these equations are different: Eq. \eq{eq:sc03} is derived under the assumption of path stretching, $t \sim c R$, in Euclidean space above a convex domain, whereas Eq. \eq{eq:wkb16} arises from a WKB-type approach in a Hyperbolic bounded Poincar\'{e} domain without an explicitly implied stretching assumption. Thus, our analysis suggests that the stretching of diffusive paths above a convex void in $\mathbb{R}^2$ may be interpreted, at the level of the leading asymptotic scaling, as a WKB-type approximation for diffusion in $\mathbb{H}^2$ near an impermeable boundary.

\section{Conclusion}
\label{sect:05}

In this paper, we have proposed arguments in support of the conjecture that diffusion in the Poincar\'{e} upper half-plane $\mathbb{H}^2$ can serve as a unified geometric and probabilistic framework linking BKT criticality, Lifshitz-tail large-deviation statistics, and KPZ-type scaling for fluctuations. 

The logic connecting the different sections of our paper is as follows. First we discussed the emergence of the BKT behavior within the RG approach. Reformulating the RG flow in terms of a stationary solution of the quasi-classical eikonal-like action which captures large-deviation behavior, we provided arguments for the asymptotic emergence of 1D Lifshitz tail. This sets the connection between BKT and LT. The link between LT and KPZ is captured in the Laplace (Mellin) transform -- see Eqs. \eq{eq:lt} and \eq{eq:sc04}. More refined discussion of emergence of KPZ-type behavior in contexts of Euclidean and Hyperbolic geometries is provided in Section \ref{sect:04}. Two appendices include supplementary materials: Appendix \ref{sect:app1} explains why we suggest that the observed 1/3-scaling is related to KPZ-type statistics, while Appendix \ref{sect:app2} provides derivation of Airy-type distribution for \eq{eq:sc03}.

One may ask why $\mathbb{H}^2$ is a natural candidate for such a unification. We highlight two key features of $\mathbb{H}^2$, discussed throughout the paper, that make it particularly well suited to this purpose:

\begin{enumerate}
\item Saddle geometry and BKT flow. When locally embedded in a three-dimensional Euclidean space, the surface of constant negative curvature, $\mathbb{H}^2$, gives rise to a saddle-like geometry. In the vicinity of the saddle point, the two-parameter RG flow exhibits one stable and one unstable direction, analogous to the flow structure of the XY model. This provides a natural geometric setting for BKT-type critical behavior and the associated divergence of the correlation length at the transition. We discuss this question in Section \ref{sect:02}.

\item Large deviations and Lifshitz tails (LT). Hyperbolic geometry provides a particularly natural setting in which large-deviation behavior can be realized: its exponential volume growth and radial symmetry translate rare fluctuations into exponentially suppressed probabilities, thereby providing the basic mechanism for the emergence of LT asymptotics near the spectral boundary, with a characteristic signature of one-dimensional Anderson localization. The corresponding description, based on an instanton-like interpretation of the RG equation, is proposed in Section \ref{sect:03}. It should be emphasized that, in contrast to the conventional mechanism of LT formation, where the LT arises from atypical fluctuations of a random walk in an external disordered potential, in our case the LT originates from atypical, curvature-induced fluctuations without external disorder.
\end{enumerate}

In Section \ref{sect:04}, we consider the statistics of 2D symmetric random walks above an impermeable disc in the Euclidean plane and demonstrate the emergence of anomalous KPZ--type scaling of fluctuations of stretched paths for various boundary conditions. Our analysis involves numeric consideration supplemented by scaling estimates and is supported by more elaborated analytical consideration presented in Appendix \ref{sect:app2}. The spectral problem for diffusion in radial geometry in Euclidean plane admits a mapping under appropriate change of variables onto the spectral problem in $\mathbb{H}^2$. Considering the last problem within the WKB approach (instead of solving the problem explicitly), we pursue two goals: 
\begin{enumerate}
\item Solving the spectral problem in $\mathbb{H}^2$ within the WKB approximation, we derive the Airy-type asymptotics of the corresponding distribution function in the vicinity of the boundary. In particular, we demonstrate that the fluctuations of paths near the boundary share the KPZ-like scaling -- see \eq{eq:wkb15} and \eq{eq:wkb19}.  
\item Comparing the WKB solution of the aforementioned spectral problem in $\mathbb{H}^2$ with its Euclidean counterpart for stretched random paths above the impermeable disc in the radial geometry, we find a similar Airy-type distribution and the KPZ-like scaling -- as given by \eq{eq:sc03}. This correspondence allows us to conjecture that the WKB approximation in $\mathbb{H}^2$ and path stretching above the hard disc in $\mathbb{R}^2$ are concentrated on similar rare-event statistical measures.
\end{enumerate}

We are fully aware that the present work has limitations and that several aspects of the proposed picture require further justification and investigation. Our aim, however, is not to claim that we constructed a complete theory, but rather to draw attention to a number of common structural features shared by seemingly different physical systems. We regard the present work as a first modest step toward identifying and exploring these connections, and hope that it will motivate further investigation of the underlying geometric and probabilistic principles.

\begin{acknowledgments}

We are grateful to Alexander Gorsky and Alexander Valov for many valuable discussions at various stages of this work. SN acknowledges the stimulating scientific environment at BIMSA (China), where part of this work was carried out. We also thank the anonymous referees for their constructive and insightful comments, which helped us improve the clarity and coherence of the manuscript.

\end{acknowledgments}

\begin{appendix}

\section{Burgers Equation and KPZ}
\label{sect:app1}

Here we provide arguments in support of our claim that a fluctuating random walk on top of a disc belongs to the family of "KPZ--like models". Consider the velocity field $u(x,t)$ governed by the stochastic Burgers equation with a small diffusivity $\nu$ and a space-time white noise source $\eta(x,t)$:
\be
\partial_t u + u\, \partial_x u = \nu \partial^2_{xx}u + \sqrt{2D}\,\partial_x \eta
\label{eq:burgers}
\ee
Define a surface height profile field $h(x,t)$ by the relation $u(x,t) = -\partial_x h(x,t)$. Integrating the Burgers equation over the spatial coordinate $x$ transforms the velocity dynamics into the canonical KPZ equation:
\be
\partial_t h = \nu \partial^2_{xx} h - \tfrac{1}{2}\left(\partial_x h\right)^2 + \sqrt{2D}\,\eta(x,t)
\ee

The unperturbed inviscid Burgers (Hopf) equation, which corresponds to $\nu=0$, $D=0$ in \eqref{eq:burgers}, reads:
\be
\partial_t u_0 + u_0\, \partial_x u_0 = 0
\label{eq:hopf}
\ee
The solution of \eq{eq:hopf} is obtained by integrating along the characteristics \cite{bec-iturriaga-khanin-2002}. For an initial boundary profile satisfying the implicit algebraic relationship $t u_0^2 - x u_0 + 1 = 0$, two possible branches are available. Selecting the physical branch that remains stable under short-time evolution gives the explicit square-root velocity field:
\be
u_0(x,t) = \frac{x - \sqrt{x^2 - 4t}}{2t}
\label{eq:sol-burg}
\ee
From a geometric point of view, \eq{eq:sol-burg} is nothing but the boundary of the disc in the Spohn--Ferrari setting. Let us point out that \eq{eq:sol-burg} exhibits a vertical gradient blow-up ($\partial_x u_0 \to \infty$) when the square root vanishes, defining the critical shock at the edge $x_c(t) = 2t^{1/2}$. Evaluating the velocity field at this critical edge, we get $u_c(t) = u_0(x_c, t) = t^{-1/2}$.

Let us analyze the localized profile of the trajectory at the edge $x_s(t)$ in the joint weak-diffusivity and zero-noise limit ($0<\nu\ll 1$, $D=0$), which means that we allow only for small fluctuations of the boundary around the equilibrium circle \cite{nowak, bulycheva}. Consider the following multiscale ansatz:
\be
x = x_s(t) + \nu^{2/3} y, \quad u(x,t) = u_0(x_c,t) + \nu^{1/3} w(y,t)
\ee
where $y$ is the rescaled and shifted $x$-variable, and $w(y,t)$ is the rescaled and shifted velocity field $u(x,t)$. In the vanishing viscosity limit ($\nu \to 0$), we arrive at the equation:
\be
w_{yy} = w w_y - \frac{1}{2}t^{-3/2}
\label{eq:w(y,t)}
\ee
The function $w(y,t)$ is considered in the quasi-stationary limit around the critical edge, where time derivatives are subdominant compared to boundary layer spatial gradients. Taking into account that $w w_y = \frac{\partial}{\partial y}\left(\frac{1}{2}w^2\right)$, we integrate \eq{eq:w(y,t)} with respect to $y$. Setting the far-field boundary integration constant to zero (which matches the inviscid root $u_0(x,t)$), yields the non-autonomous Riccati equation:
\be
w_y = \frac{1}{2}w^2 - \frac{y}{2t^{3/2}}
\label{eq:ric}
\ee

To linearize the Riccati equation, we express the localized velocity profile $w(y,t)$ as the logarithmic derivative of an auxiliary field $\phi(y,t)$:
\be
w(y,t) = -2 \partial_y \ln \phi(y,t) = -2 \frac{\phi_y}{\phi}
\label{eq:lin-ric}
\ee
Differentiating \eq{eq:lin-ric} with respect to $y$ results in:
\be
w_y = -2 \frac{\phi_{yy}}{\phi} + 2 \left( \frac{\phi_y}{\phi} \right)^2
\ee
Substituting both $w$ and $w_y$ back into the Riccati equation gives:
\be
-2 \frac{\phi_{yy}}{\phi} + 2 \left( \frac{\phi_y}{\phi} \right)^2 = 2 \left( \frac{\phi_y}{\phi} \right)^2 - \frac{y}{2t^{3/2}}
\ee
The nonlinear squared terms cancel out. Multiplying through by $-\frac{1}{2}\phi$ provides the linear second-order ordinary differential equation:
\be
\phi_{yy} - \frac{y}{4t^{3/2}} \phi = 0
\ee
Introducing the scaled coordinate $\zeta = (4t^{3/2})^{-1/3}y$, we arrive at the classical Airy differential equation:
\be
\phi(y,t) = \text{Ai}\left( \frac{y}{(4t^{3/2})^{1/3}} \right)
\ee
Thus, the smooth regularized solution inside the localized layer is:
\be
w(y,t) = -2 \frac{\partial}{\partial y} \ln \left(\text{Ai}\left( \frac{y}{(4t^{3/2})^{1/3}} \right) \right) = -\frac{2^{1/3}}{\sqrt{t}} \frac{\text{Ai}'\left( \frac{y}{2^{2/3}\sqrt{t}} \right)}{\text{Ai}\left( \frac{y}{2^{2/3}\sqrt{t}} \right)}
\label{eq:airy2}
\ee

Let us demonstrate that the solution \eq{eq:airy2} belongs to the family of solutions of a Painlevé II equation. The standard Painlevé II equation is a non-linear second-order ordinary differential equation defined by:
\be
u''(s) = 2u(s)^3 + s u(s) + a
\label{eq:pain}
\ee
where $a$ is a complex parameter. Generically for $a \neq 0$, \eq{eq:pain} defines transcendental functions. However, when $a = \pm \frac{1}{2}$, the equation admits a special non-linear first-order reduction. Consider a function $w(s)$ satisfying the non-autonomous Riccati equation derived from our noiseless crossover layer \eq{eq:ric} (in dimensionless units):
\be
w' = \frac{1}{2}w^2 - \frac{1}{2}s
\label{eq:riccati}
\ee
Let us define a field $u(s) = -\frac{1}{2}w(s)$. Differentiating $u(s)$ twice with respect to $s$ yields:
\be
\begin{aligned}
u' &= -\frac{1}{2}w' = -\frac{1}{4}w^2 + \frac{1}{4}s = -u^2 + \frac{1}{4}s \\
u'' &= -2uu' + \frac{1}{4} = -2u\left(-u^2 + \frac{1}{4}s\right) + \frac{1}{4} = 2u^3 - \frac{1}{2}su + \frac{1}{4}
\end{aligned}
\ee
By rescaling the independent and dependent variables according to $s=-2^{-1/3}x$ and $u(s)=2^{-1/3}\tilde u(x)$, the resulting second-order equation becomes
\be
\tilde{u}''(x)
=
2\tilde{u}(x)^3+x\tilde{u}(x)-\frac12.
\ee
This demonstrates that the Burgers equation with small diffusivity but without external noise $\eta(x,t)$ is a linearizable solution of the Painlevé II equation for $a = -1/2$. Recall that the solution to \eq{eq:riccati} is given by the logarithmic derivative of the Airy function:
\be
w(s) = -2\partial_s \ln \text{Ai}\left(2^{-1/3}s\right)
\ee
Introducing external fluctuations $\eta(x,t)$ alters the deterministic shock structure. While the mean localized velocity profile in the noiseless limit is governed by the special linearizable case $a = -1/2$ of Painlevé II (yielding the Airy profile), the cumulative distribution function of height fluctuations in the presence of white noise is governed by the Hastings--McLeod solution \cite{hastings-mcleod-1980} corresponding to $a = 0$:
\be
\bar{u}''(s) = 2\bar{u}^3(s) + s \bar{u}(s)
\ee
where $\bar{u}(s) \sim \text{Ai}(s)$ as $s \to \infty$. Thus, noise drives the system from a deterministic Riccati profile into the Tracy--Widom universality class while preserving the underlying spatial fluctuation scaling of KPZ systems.

\section{Statistics of inflated Brownian paths above an impermeable disc in 2D}
\label{sect:app2}

We compute the first-return probability of stretched paths above a disc with Dirichlet boundary conditions, following the line of reasoning proposed in \cite{nech-val} and \cite{fedotov-nech}. We focus on the double-scaling regime $t \to \infty$ and $R \to \infty$ with $t/R=c=\mathrm{const}$. The step length is set to 1 as everywhere in the text. To access this large-deviation regime analytically, we temporarily treat $t$ and $R$ as independent parameters and control the stretching by introducing an external transverse magnetic field. Its strength is then fixed self-consistently so that the corresponding Larmor orbit has radius $R_c$.

Consider an ensemble of inflated two-dimensional $N$-step random loops of total length $t$ surrounding an impermeable disc of radius $R_d$. The algebraic area $A$ is driven to the atypical strongly inflated circular shape $A_c=\pi R_c^2$ with $R_c=c R_d$ ($c\geq 1$). Introducing the Legendre--conjugate variable $q$, which controls the area, we have the following path-integral representation \cite{khandekar}
\be
P(r,t) = \frac{1}{2\pi}\int_{-\infty}^{\infty}dq e^{-iq A_d}\int{\cal D}{\mathbf{r}(t')} e^{-\int_0^t L(\mathbf{r}, \dot{\mathbf{r}},t') dt'}
\label{eq:c01}
\ee
with
\be
L(\mathbf{r}, \dot{\mathbf{r}},t) = \left(\dot{x}^2+\dot{y}^2\right)- \frac{i q }{2}\left(x \dot{y}-y\dot{x} \right)
\label{eq:c02}
\ee
This is equivalent to the action of a charged particle in a constant magnetic field, where
\be
\mathbf{A}(\mathbf{r}(t)) = \mathbf{B} \times \mathbf{r}(t), \quad \mathbf{B} = (0,0,B)
\label{eq:c03}
\ee

The classical trajectories are Larmor orbits,
\be
\begin{cases}
x(t) = R \cos(\omega t) \\ y(t) = R \sin(\omega t)
\end{cases}; \quad \omega = \frac{|q| B}{2}; \quad R = \frac{2}{|q| B}
\label{eq:c04}
\ee
Choosing $B=2$ and requiring the orbit radius to be $R_c$ (which is slightly bigger than the disc radius, $R_d$), gives us
\be
q = \frac{1}{R_c}
\label{eq:c05}
\ee
Thus, the magnetic field provides an effective description of the pre-inflated Brownian loops. The corresponding diffusion equation follows from the Hamiltonian $H = \frac{1}{4}\left(\hat{p}+ iq A\right)^2$, where $\hat{p} = \nabla_{r,\varphi}$. In polar coordinates the diffusion equation reads,
\be
\frac{1}{4}\left(\frac{1}{r}\frac{\partial}{\partial r}\left(r \frac{\partial}{\partial r}\right) + \frac{1}{r^2}\frac{\partial^2}{\partial\varphi^2} + 2iq \frac{\partial}{\partial \varphi} - q^2 r^2\right)P(r,\varphi,t)=\frac{\partial}{\partial t}P(r,\varphi,t)
\label{eq:c07}
\ee
Imposing the Dirichlet condition at $r=R_d$, separating variables as $P(r,\varphi,t)=e^{-\lambda t}W(r)e^{im\varphi}$ and performing substitution $W(r)=U(r)r^{-1/2}$, the radial problem becomes
\be
\begin{cases}
\disp \frac{d^2U(r)}{dr^2}- \left(q^2r^2-\frac{1}{4r^2}-4\lambda + 2qm \right) U(r) = 0 \medskip \
\disp U(r=R_d,t) = U(r\to\infty,t)=0
\end{cases}
\label{eq:c08}
\ee
For $r=R_d+\rho$ with $0<\frac{\rho}{R_d}\ll1$, the radial equation reduces to
\be
\frac{d^2U(\rho)}{d \rho^2}- \alpha_1 U(\rho) - \alpha_2 \rho, \quad U(\rho) = 0
\label{eq:c09}
\ee
where
\be
\alpha_1=q^2R_d^2-\frac{1}{4R_d^2}-\frac{4}{a^2}\lambda+2q m, \qquad \alpha_2=2q^2R_d+\frac{1}{2R_d^3}
\label{eq:c10}
\ee
The decaying solutions satisfying the Dirichlet condition are expressed through Airy functions,
\be
U_j(\rho)\propto \mathrm{Ai}\left(\alpha_2^{1/3} \rho+\alpha_1 \alpha_2^{-2/3}\right),
\label{eq:c11}
\ee
where $\alpha_1 \alpha_2^{-2/3}=a_j\le 0$ are the zeros of the Airy function, i.e. $\mathrm{Ai}(a_j)=0$ for $j=1,2,...$ and we consider the first zero $a_1\approx-2.338$ only. Using $q=\frac{1}{R_c}$ and taking the large--$R_d$ limit, we obtain
\be
\begin{cases}
\disp \alpha_2\approx 2q^2 R_d=\frac{2 R_d}{R_c^2}; \\ \disp \lambda_{jm} \approx \frac{1}{4}\left(q^2 R_d^2 +2 q m + |a_j|(2q^2 R_d)^{2/3}\right) = \frac{1}{4}\left(\frac{R_d}{R_c} \right)^2 + \frac{m a}{2R_c} + \frac{|a_j|}{2^{4/3}}\left(\frac{R_d}{R_c^2} \right)^{2/3}
\end{cases}
\label{eq:c12}
\ee
Setting $R_c=cR_d$ and using $N=2\pi R_c$, \eq{eq:c12} can be rewritten as follows
\be
\begin{cases}
\disp \alpha_2\approx \frac{1}{\pi a^3 c t}; \medskip \\ \disp \lambda_{jm} \approx \frac{1}{4c^2}  + \frac{|a_j|}{4 (\pi c t)^{2/3}}+ \frac{m}{4\pi t}
\end{cases}
\label{eq:c13}
\ee

The probability distribution $P(\rho,\rho',N)$ can then be expanded in the radial eigenfunctions,
\be
P(\rho,\rho',t)=\sum\limits_{j=1}^{\infty}\sum_{m=1}^{\infty} \frac{C_j}{R_d} e^{-\lambda_{jm} t} U_j(\rho) U_j^*(\rho'),
\label{eq:c13a}
\ee
where $(...)^*$ denotes complex conjugation. Restricting to the single-winding sector, $m=1$, and retaining the leading finite-size correction, the expression for the Brownian bridge partition function gives us
\be
P(\rho,\rho'=\rho,t)=\frac{C}{R_d} e^{-\frac{t}{4c^2} - \frac{|a_1|t^{1/3}}{4 (\pi c)^{2/3}}} \mathrm{Ai}^2\left(\frac{2^{1/3}\rho}{(\pi c t)^{1/3}}-|a_1|\right)
\label{eq:c14}
\ee
where the constant $C$ is fixed by the initial and normalization conditions. Equation \eq{eq:c14} yields two important consequences:
\begin{enumerate}
\item The typical span $\Delta$ of a random walk of length $t$ around a disc of radius $R_d$ in the stretched regime $N=2\pi R_c=2\pi cR_d$ is
\be
\Delta \approx \frac{|a_1| \pi^{1/3}}{2^{1/3}}c^{1/3} t^{1/3}
\label{eq:c14a}
\ee
which up to numerical constants reproduces \eq{eq:sc03} and \eq{eq:wkb19}.
\item In the ground-state-dominance regime, $t\gg 1$, the distribution has the scaling form
\be
P(t) \approx e^{-\frac{t}{4c^2} - \frac{|a_1|t^{1/3}}{4 (\pi c)^{2/3}}}
\label{eq:c15}
\ee
\end{enumerate}
The scaling with $t$ and $c$ in \eq{eq:c14a}-\eq{eq:c15} is consistent with the arguments provided after \eq{eq:sc03}. The additional linear term in $t$ in the exponent of \eq{eq:c15} emerges due to the following. In \eq{eq:sc03}, stretching emerges dynamically from the double-scaling limit $t\gg 1$, $R\gg 1$, with $\frac{t}{R}=c=\mathrm{const}$. In contrast, in the derivation of \eq{eq:c15}, the trajectories are pre-inflated by the external magnetic field, which effectively confines the spatial extent of the paths. This confinement gives rise to an additional exponential factor, analogous to the effect of a bounded domain.

\end{appendix}

\bibliography{biblio-3faces}

\end{document}